\documentclass[compsoc,journal]{mytran}

\RequirePackage[compress]{cite}
\usepackage[utf8]{inputenc}
\usepackage[T1]{fontenc}
\usepackage[english]{babel}
\usepackage{textcomp}
\usepackage{url}
\usepackage{graphicx}
\usepackage{float}
\usepackage{multirow}
\usepackage{booktabs}
\usepackage{verbatim}

\usepackage{multicol}
\usepackage{xcolor}

\usepackage{amsmath, amsfonts, mathtools, amsthm,physics,amssymb}

\usepackage{subfig}
\usepackage{adjustbox}

\newcommand{\bmu}{\boldsymbol{\mu}}

\newcommand{\citep}{\cite}
\newcommand{\citet}{\cite}

\usepackage{color,array,verbatim,algpseudocode,bm,algorithm,accents}
\definecolor{menucolor}{rgb}{0.1,0.52,0.47}
\definecolor{urlcolor}{rgb}{0.85,0.37,0.01}
\definecolor{runcolor}{rgb}{0.46,0.44,0.701}
\definecolor{filecolor}{rgb}{0.2,0.5,0.01}
\definecolor{linkcolor}{rgb}{0.12,0.47,0.70}
\definecolor{citecolor}{rgb}{0.55,0.36,0.01}
\definecolor{anchorcolor}{rgb}{0.4,0.4,0.4}
\usepackage[colorlinks=true, urlcolor=urlcolor, linkcolor=linkcolor,citecolor=citecolor,filecolor=filecolor, anchorcolor=anchorcolor,menucolor=menucolor]{hyperref}
\usepackage{nameref}
\usepackage{zref-xr}

\makeatletter
\newcommand\VeryLarge{\@setfontsize\Huge{16}{16}}
\newcommand\smaller{\@setfontsize\small{5}{5}}
\makeatother   

\def\subparagraph{} 
\usepackage{titlesec}
\titlespacing*{\section}{0pt}{*1}{*1}
\titlespacing{\subsection}{0pt}{*1}{*1}
\titlespacing{\subsubsection}{0pt}{*1}{*1}

\let\LambdaOLD\Lambda
\renewcommand{\Lambda}{{\bm\LambdaOLD}}
\let\Psiold\Psi
\renewcommand{\Psi}{{\bm\Psiold}}
\let\Sigmaold\Sigma
\renewcommand{\Sigma}{{\bm\Sigmaold}}

\newcommand{\lm}{\bm{\Lambda}}

\newcommand{\si}{\bm{\Sigma}}
\newcommand{\psii}{\bm{\Psi}}

\newcommand{\Mu}{\bm{\mu}}
\newcommand{\eps}{\bm{\epsilon}}

\newcommand{\yy}{\bm{y}}

\newcommand{\xx}{\bm{x}}

\newtheorem{lemma}{Lemma}
\newtheorem{result}{Result}

\titleformat{\paragraph}[runin]{\normalfont\bfseries}{\theparagraph}{0.5em}{}[:]

\IEEEaftertitletext{\vspace{-3\baselineskip}} 

\makeatletter
\renewenvironment{proof}[1][\proofname]{\par
  \vspace{-\topsep}
  \pushQED{\qed}%
  \normalfont
  \topsep1pt \partopsep1pt 
  \trivlist
  \item[\hskip\labelsep
        \itshape
    #1\@addpunct{.}]\ignorespaces
}{%
  \popQED\endtrivlist\@endpefalse
  \addvspace{1pt plus 1pt} 
}
\makeatother

\begin{document}
\title{\Large A Hybrid Mixture Approach for Clustering and Characterizing Cancer Data}
\author{
  \vspace{-0.15in}
  Kazeem Kareem and Fan Dai\\
  \vspace{-0.4in}
  \thanks{K. A. Kareem and F. Dai are with the Department of Mathematical Sciences, Michigan Technological University, Houghton, MI 49931, USA.}
    
}
\maketitle
\begin{abstract}
Model-based clustering is widely used for identifying and distinguishing types of diseases. However, modern biomedical data coming with high dimensions make it challenging to perform the model estimation in traditional cluster analysis. The incorporation of factor analyzer into the mixture model provides a way to characterize the large set of data features, but the current estimation method is computationally impractical for massive data due to the intrinsic slow convergence of the embedded algorithms, and the incapability to vary the size of the factor analyzers, preventing the implementation of a generalized mixture of factor analyzers and further characterization of the data clusters. We propose a hybrid matrix-free computational scheme to efficiently estimate the clusters and model parameters based on a Gaussian mixture along with generalized factor analyzers to summarize the large number of variables using a small set of underlying factors. Our approach outperforms the existing method with faster convergence while maintaining high clustering accuracy. Our algorithms are applied to accurately identify and distinguish types of breast cancer based on large tumor samples, and to provide a generalized characterization for subtypes of lymphoma using massive gene records.


\end{abstract}

\section{Introduction} \label {introduction}
 Cluster analysis has found widespread applications in biological and medical studies, for example, grouping tumor samples with a similar molecular profile \citep{otuetal2005}, defining signature gene expression profiles from isolated populations of muscle cells \citep{choietal2020}, and analyzing gene types associated with diseases \citep{eisenetal1998, Selinskietal2008}. Commonly used techniques include the nonparametric and model-based clustering, where the nonparametric methods, such as hierarchical clustering \citep{everitt1974}, \textit{k}-means \citep{macqueen1967},
fuzzy \textit{c}-means \citep{bezdek1981}, mean shift \citep{comaniciuetal2002}, and spectral clustering \citep{ngetal2002}, rely on similarities or distance measures between data points, making them versatile but often sensitive to the choice of hyperparameters, and the computational complexity increases rapidly as the sample size grows.
 
 In contrast, model-based clustering \citep{mclachlanpeel2000,mclachlanetal2003,melnykovetal2012} assumes that the data are generated from a probabilistic model with specific underlying distributions for individual groups. One prominent example is the Gaussian mixture model (GMM) \citep{mclachlanpeel2000}, where data are from a mixture of Gaussian distributions. Estimating the mixture model involves maximization of the data likelihood which are normally done via the Expectation-Maximization (EM) algorithms \citep{dempsteretal1977,mclachlanpeel2000,tippingBishop1999} where the parameter estimates can be easily obtained from the likelihood function of the augmented data with the unobserved group indicators. However, the EM-type algorithms suffer from slow convergence and local maximization, which become more severe when the data dimension increases. 
 
 Given the challenge from the high-dimensional problems, a mixture of factor analyzers (MFA) \citep{ghahramaniHinton1996, andrewsetal2011, mcnicholas2008} is built upon the GMM by specifying a lower-dimensional representation of the covariance matrix for each Gaussian group. Consequently, the MFA leverages the mixture model to identify local structures tailored to individual clusters, thus capturing both global trends and local variations \citet{mclachlanpeel2000, ghahramaniHinton1996}. However, the MFA still faces several issues regarding estimating the model parameters. Current methods for obtaining the maximum likelihood estimates \citep{mclachlanpeel2000,mclachlanetal2003} employ EM algorithms for both the mixture components and factor analyzers, aggravating the inefficiency inherent in the EM iterations, and making it computationally impractical for data where the number of features is notably large and exceeds the sample size \citep{daietal2020}, which, however, occurs frequently for biomedical datasets. On the other hand, the existing MFA algorithms usually assume a common number of factors across all the clusters, preventing the use of different factor sizes to characterize the identified groups.

We propose a hybrid approach that adopts the EM framework for mixtures but a matrix-free computational scheme for the factor analyzers using the profile likelihood method introduced by \cite{daietal2020} to gain computational efficiency. We test our method and compare it to current algorithm via simulated datasets, which shows that the proposed approach achieves a higher speed of convergence without sacrificing the clustering and estimation accuracy as demonstrated in Section \ref{methodology}. We further extend our method to a generalized MFA where clusters are allowed to have different numbers of factors, providing a more flexible way to  characterize the dispersion of data. In Section \ref{application}, we apply our approaches to cluster and characterize the breast cancer and lymphoma data, revealing distinguishable grouping patterns for the correctly identified subtypes of diseases. We
conclude with a summary of the contributions of our work and discuss further extensions.

\section{Methods and Algorithms} 
\label{methodology}
We first discuss the MFA model and estimation methods. We provide descriptions of current MFA algorithms, our approaches and algorithms, which are further illustrated via simulation studies. 
\subsection{Background and preliminaries}
\label{prelim}
 \subsubsection{Gaussian mixture model}
\label{EM:mfae}
Suppose a $p$-dimensional random vector $\yy$ comes from the GMM. Then, its density function is given by,  
\begin{equation} 
f(\yy; \bm{\theta})= \sum_{k=1}^K  \omega_k f_{\mathcal{N}_p}(\yy;\Mu_k, \Sigma_k ), \label{gaussianmixture}
\end{equation}
where $\omega_k >0$ for each $k\in \{1, \ldots ,K\}$ with $\sum_{k=1}^K  \omega_k=1$, and $\yy$ is said to belong to the $k$th component with probability $\omega_k$. $\Mu_k$ and $\Sigma_k$ represent the mean and covariance parameters for the $k$th Gaussian component, and $\bm{\theta}$ denotes the entire parameter space.

The EM algorithm, outlined in \citep{dempsteretal1977,redneretal1984,bishop2006}, is the most commonly used technique for estimating the parameters of GMM, where an unobserved group indicator $z_{ik}$ is assumed for the $i$th data point $\yy_i$, $i=1,2,\ldots,n$ so that $z_{ik}$ equals 1 if $\yy_i$ is assigned to the $k$th cluster and 0 otherwise. The EM algorithm then constructs a complete data including both the observed $\yy_i$ and the latent $z_{ik}$ to obtain a complete data log-likelihood function ($Q$-function) for easy optimization, and iterates between the expectation (E) step that computes the conditional expectations of the unobserved quantities and the maximization (M) step that solves for parameter estimates by optimizing the $Q$-function with the expectations until results converge. For GMM, the EM iteration \citep{mclachlanpeel2000} is given below.

\textbf{E Step.}
We compute the expectation of $z_{ik}$ given observed data as 
\begin{equation}
\label{gmm:estep}
\begin{split}
\gamma_{ik} = \mathbb{E}[\mathbf{I}(z_{ik}=1|\yy_i)]=\frac{\omega_kf_{\mathcal{N}_p}(\yy_i;\Mu_k,\si_k)}{\sum_{j=1}^K \omega_j f_{\mathcal{N}_p}(\yy_i;\Mu_j,\si_j)},
\end{split}
\end{equation}
where $\mathbf{I}(\cdot)$ denotes the indicator function.

\textbf{M Step.} Then, for each cluster, we update the parameter estimates as follows.
\begin{equation}
\label{gmm:mstep}
\begin{split}
&\hat{\omega}_k = n^{-1}\sum_{i=1}^n \gamma_{ik}  \\
&\hat{\Mu}_k =  \frac{\sum_{i=1}^n \gamma_{ik}\yy_i}{ \sum_{i=1}^n \gamma_{ik}} \\
 &\hat{\Sigma}_k =  \frac{\sum_{i=1}^n \gamma_{ik}(\yy_i- \hat{\Mu}_k)(\yy_i- \hat{\Mu}_k)^\top
 }{ \sum_{i=1}^n \gamma_{ik}}.
 \end{split}
\end{equation}

  \subsubsection{Mixture of factor analyzers}
  \label{mfa-aecm}
The MFA incorporates a factor model to each mixture component of the GMM. Consequently, the data points from the MFA can be represented as
\begin{equation}
 \yy_i|(z_{ik}=1) = \Mu_k+ \Lambda_k\xx_{ik}+ \eps_{ik},   
\end{equation}
where $\xx_{ik} \sim \mathcal{N}_{q_k}(\bm{0}, \bm{I}_{q_k} )$ represents the $q_k$ latent factors, independent of $\eps_{ik} \sim \mathcal{N}_{p}(\bm{0}, \psii_k )$. $\Lambda_k$ is a $p\times q_k$ loading matrix of rank $q_k<\min(n,p)$ and $(p-q_k)^2>p+q_k$, which explains the common variances shared by all the $p$ variables for the $k$th group, and $\Psi_k$ is a $p\times p$ diagonal matrix of unique variances for the $k$th group. By the setting above, we obtain a lower-dimensional representation of the $k$th covariance matrix as
\begin{equation} \label{fa.assumption}
   \Sigma_k = \Lambda_k\Lambda_k^{\top}+\Psi_k .
\end{equation}

\citet{mclachlanetal2003} proposes an alternating expectation-conditional maximization (AECM) algorithm with a common number of factors $q_1 = q_2=\ldots=q_K=q$ to estimate the MFA parameters. The method essentially combines the EM algorithm for clustering data into components as introduced in Section \ref{EM:mfae} and an EM algorithm that performs local factor analysis on each of the components where the data is augmented with the underlying factors $\xx_{ik}$. The algorithm is implemented in the R package \texttt{EMMIXmfa} and is referred to as EMMIX. While EMMIX can reduce the data dimension through the factor analyzers, it still suffers slow convergence due to the double EM iterations, especially for data with $n<p$, making it challenging for the algorithm to scale. Hence, we develop a hybrid expectation-conditional maximization (ECM) framework that embeds matrix-free computations for factor models (with a common $q$) in the EM for mixture components in order to reduce the computational cost and memory usage. 

\subsection{A hybrid ECM algorithm for MFA}

As per Section \ref{EM:mfae}, we construct the E step as described in \eqref{gmm:estep}. Next, in the conditional
maximization (CM) step, we firstly compute $\hat{\omega}_k$, $\hat{\Mu}_k$ and $\hat{\Sigma}_k$ according to \eqref{gmm:mstep}, then, given $\hat{\Sigma}_k$, we jointly update the two covariance parameters from the factor model, $\Lambda_k$ and $\Psi_k$ by adapting the profile likelihood method developed by \cite{daietal2020}. Specifically, $\Lambda_k$ can be profiled out from the $Q$-function following the result below.
\begin{result}\label{meth-profile}
  For a positive-definite diagonal matrix $\Psi_k$, let $\theta_{1,k} \geq
  \theta_{2,k} \geq \cdots \geq \theta_{q,k}$ be the $q$ largest eigenvalues
  of ${\bm{G}_k} = \Psi_k^{-1/2}\hat{\Sigma}_{k}\Psi_k^{-1/2}$. Let the columns of
  $\bm{V}_{k}$ store the eigenvectors corresponding to these $q$ eigenvalues. Then the $Q$-function is maximized w.r.t. $\Lambda_k$ at ${\hat\Lambda_k} = \Psi_k^{1/2}{\bm V}_{k}\bm\Delta_{k},$ where $\bm\Delta_k$ is a $q\times q$ diagonal matrix with $j$th diagonal entry $[\max(\theta_{j,k}-1,0)]^{1/2}.$ The profile $Q$-function for the $k$th group is then given by
\begin{equation}
\label{eqn:meth-profile}
\begin{split}
Q_p(\Psi_{k}) = 
c&-\frac{\hat{\omega}_{k}n}{2}\{\log\det\Psi_{k} + \Tr \Psi_{k}^{-1}\hat{\Sigma}_k
 \\
 &+\sum_{j=1}^{q_{k}}(\log\theta_{j,k} - \theta_{j,k} + 1)\}
\end{split}
\end{equation}
where $c$ is a constant independent of $\Psi_k$.
\end{result}
\begin{proof}
The proof follows the Lemma 1 in \cite{daietal2020} by adapting the profile log-likelihood function with the group-wise "sample covariance" matrix $\hat{\Sigma}_k$.
\end{proof}

In Result \ref{meth-profile}, the $q$ largest eigenvalues and the associated eigenvectors can be accurately approximated through the Lanczos algorithm \citet{baglamareichel2005} within a few iterations. Next, we optimize the profile log-likelihood function \eqref{eqn:meth-profile} w.r.t $\Psi_k$ using the limited-memory Broyden-Fletcher-Goldfarb-Shanno quasi Newton algorithm with box constraints (L-BFGS-B) \cite{byrd1995} algorithm and finally update $\Lambda_k $ using the relation $\hat{\lm}_k = \hat{\psii}_k^{1/2}\bm{V}_{k}\bm{\Delta}_k $ as defined in Result \ref{meth-profile}. Both the Lanczos and L-BFGS-B algorithms involve only matrix-vector multiplications that avoid the storage of large $p\times p$ matrices, which is learned as the matrix-free property. Our algorithm is called GMMFAD.
  
\subsection{Initialization, stopping criteria and model selection}
\label{init-stop-bic}
As mentioned in Section \ref{introduction}, to mitigate the local maximization of EM, we implement a random initialization method proposed by \citet{Biernackietal2003,Maitra2013}, where the algorithm starts with a large set of random initial values and run for a few iterations, then we select a few candidate initials with the highest data log-likelihood values, after that, the algorithm will run with the selected initials until convergence and the optimal result is determined as the one giving the highest final data log-likelihood. Besides, we also include an extra initialization from k-means clustering as the data-driven initials. The algorithm stops when there is no more significant increase in the data log-likelihood value with a tolerance level of $10^{-6}$ in practice, or when the iterations reach $500$. We determine the best number of groups and factors by the Bayesian Information Criteria (BIC) \citep{schwarz1978}, where the best model would have the lowest BIC value. 

\subsection{Generalized MFA with varying $q_k$}
Further, we propose a generalized MFA approach by allowing different numbers of factors across the clusters, which cannot be handled by EMMIX that requires a common $q$ as explained in Section \ref{mfa-aecm}. This extension facilitates the characterization of different data clusters and can be easily implemented with a modified version of GMMFAD, which is named GMMFAD-q. The generalized MFA would need an extra constraint on $q_k$ to guarantee the model estimability as follows,

\begin{lemma}
$$\max_{k\in \{1 , \ldots, K\} } q_k< p + (1-\sqrt{1+8p})/2. $$
\end{lemma}\label{lemma1}
\begin{proof}
The condition that $ \Lambda_k^{\top}\Psi_k^{-1}\Lambda_k $  be diagonal imposes $\frac{1}{2}q_k(q_k-1)$ constraints on the parameters \citep{lawleymaxwell1971}.  Hence, for each $k \in \{1, \ldots, K\}$, the number of free parameters in the factor analytic model is 
\begin{equation}
pq_k+p -\frac{1}{2}q_k(q_k-1) . 
\end{equation}
Suppose $s_k$ is  the difference between the number of parameters for $\Sigma_k$ and the number of free parameters considering the assumption \ref{fa.assumption}. Then for each $k\in \{1, \ldots K\}$, 

\begin{align}
  s_k &=  \frac{1}{2}p(p-1) - \left(pq_k + p -\frac{1}{2}q_k(q_k-1) \right)\\
      &=  \frac{1}{2}\left[ (p-q_k)^2- (p+q_k) \right]
\end{align}
This difference represents the reduction in the number of parameters for $\Sigma_k$. For this difference to be positive, each $q_k$ needs to be small enough so that

\begin{align*}
       &\frac{1}{2}\left[ (p-q_k)^2- (p+q_k) \right] > 0  \quad    \forall k \\
       \implies& \quad q_k <  p + (1-\sqrt{1+8p})/2   \quad \forall k \\
       \implies& \max_{k\in \{1 , \ldots, K\} } q_k <  p + (1-\sqrt{1+8p})/2.
\end{align*}
\end{proof}

\subsection{Comparison studies between GMMFAD and EMMIX}
\begin{figure}[!ht]
  \centering
\mbox{
\subfloat[$\bar{\omega}=0.001$]{
\includegraphics[width=0.31\linewidth]{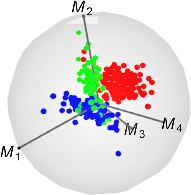}
}
}%
\mbox{
\subfloat[$\bar{\omega}=0.005$]{
\includegraphics[width=0.3\linewidth]{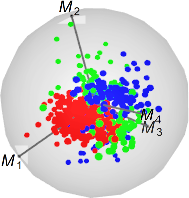}
}
}%
\mbox{
\subfloat[$\bar{\omega}=0.01$]{
\includegraphics[width=0.31\linewidth]{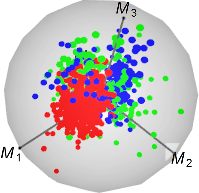}
}
}
\caption{3D displays of the MFA datasets with varying overlap rates for $n=300, p=10, K=3, q=2$. Plots were generated using the 3D radial visualization tool developed by \cite{Zhuetal2022}.}
\label{radviz:mfa}
\end{figure}

We compare the performance of GMMFAD and EMMIX via simulated datasets. The clustering complexity among all the clusters is specified by the generalized overlap rate \citep{melnykovetal2012,maitraandmelnykov2010} $\bar{\omega} = 0.001,0.005, 0.01$, where smaller $\bar{\omega}$ indicates more separations, as shown by Figure \ref{radviz:mfa}.
For $n=300, p=10$ with all combinations of $q=2,3$, $K=2,3$ and the three $\bar{\omega}$ values, we obtain the true parameter values of $\Mu_k$ and $\Lambda_k$ from standard normals, values of $\Psi_k$ from $\mathrm{Unif}(0.2,0.8)$, and the grouping probabilities $\omega_1,\omega_2,\ldots,\omega_K$ with standard normals and scale the absolute values to have a sum of 1. Then, we generate 100 grouped Gaussian datasets via the R package \texttt{MixSim} \citep{maitraandmelnykov2010} given the prefixed parameters. Each dataset is fitted using GMMFAD and EMMIX with up to $2K$ groups and up to $2q$ factors, respectively, with the same initialization method and stopping rules described in Section \ref{init-stop-bic}. All experiments were done using R \citep{R} on the same machine.

\begin{figure}[h!]
    \centering
\includegraphics[width=0.485\textwidth]{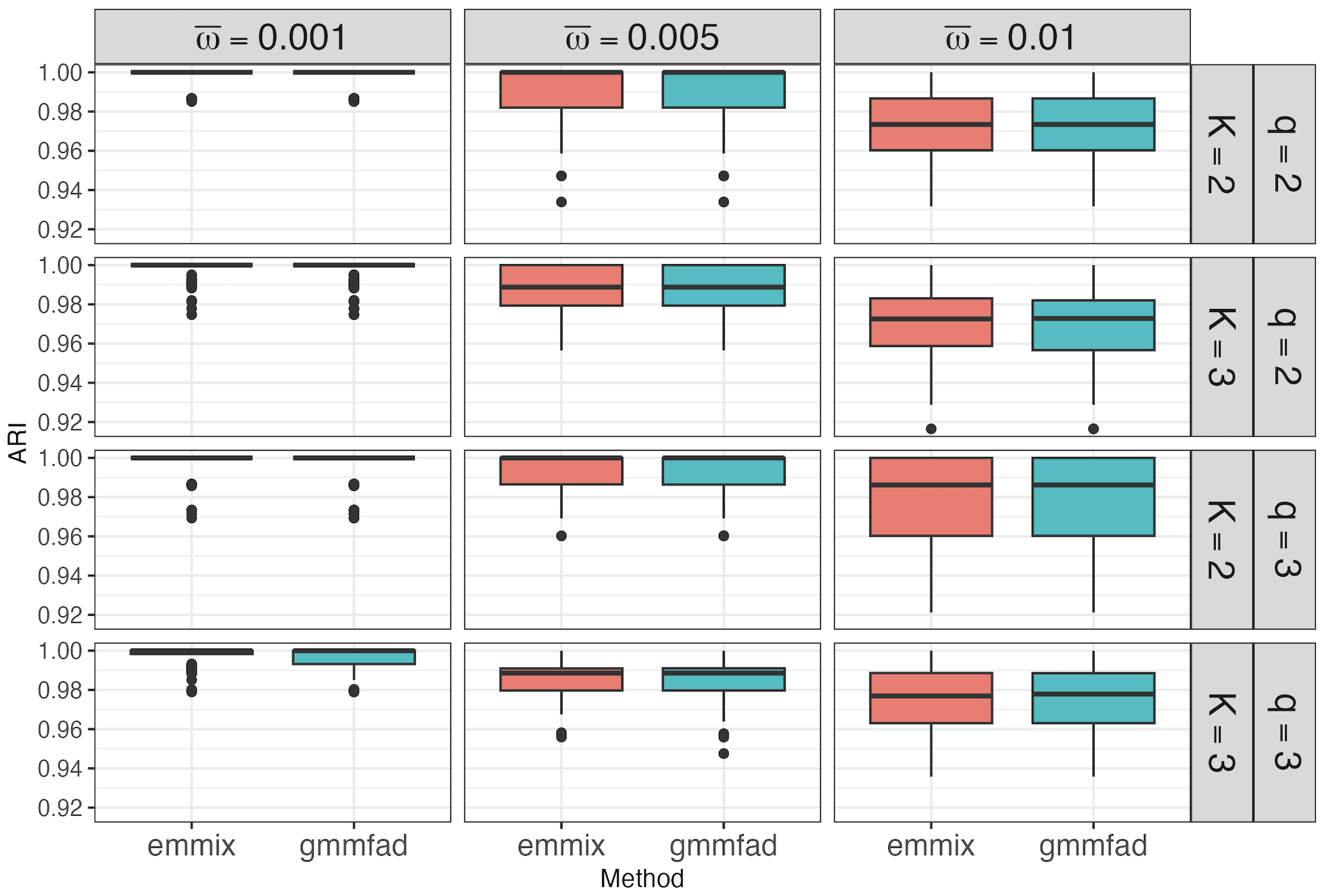} 
    \caption{Boxplots of the ARI values fitted with GMMFAD and EMMIX for $n=300,p=10$, with colors for methods.}
    \label{ari_plots}
\end{figure}

The model correctness rates, computed as the percentage of runs where the BIC chooses an optimal model with correct $K$ and $q$, are above $98\%$ for both GMMFAD and EMMIX across all the settings. Given the correct models, we evaluate the similarity between the true and estimated clusters using the adjusted rand index (ARI) \citep{melnykovetal2012}, as displayed in Figure \ref{ari_plots}. We can see that both GMMFAD and EMMIX achieved high and almost identical clustering accuracy for all the simulation runs. Similar patterns also appear for parameter estimation results, where we evaluate the accuracy of estimates compared to the true values by the relative Frobenius distance, for example, for $\Lambda_k\Lambda_k^\top$ instead of $\Lambda_k$ due to the identifiability, the relative distance is computed as $d_{\Lambda_k\Lambda_k^\top} = \|\hat{\Lambda}_k\hat{\Lambda}_k^\top - \Lambda_k\Lambda_k^\top\|_F/\|\Lambda_k\Lambda_k^\top\|_F$. GMMFAD and EMMIX reached nearly identical estimation accuracy as shown by Figure S1 
of the Supplement.

\begin{figure}[h!]
    \centering
    \mbox{
\subfloat[$n=300,p=10$]{\label{fig:speedup-p10}
\includegraphics[width=0.85\linewidth]{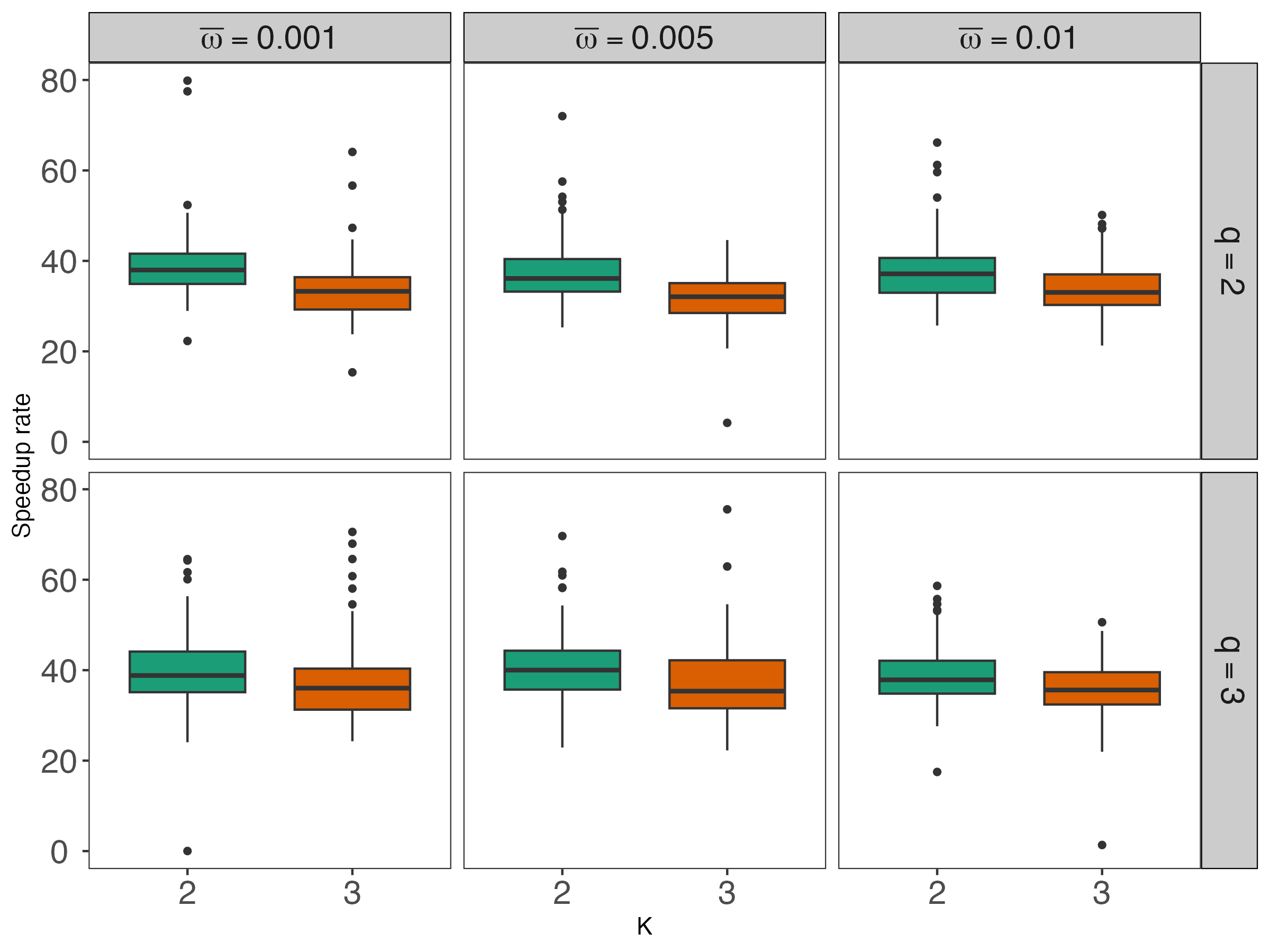}
}
}\\
\mbox{
\subfloat[$n=150,p=150$]{\label{fig:speedup-p150}
\includegraphics[width=0.85\linewidth]{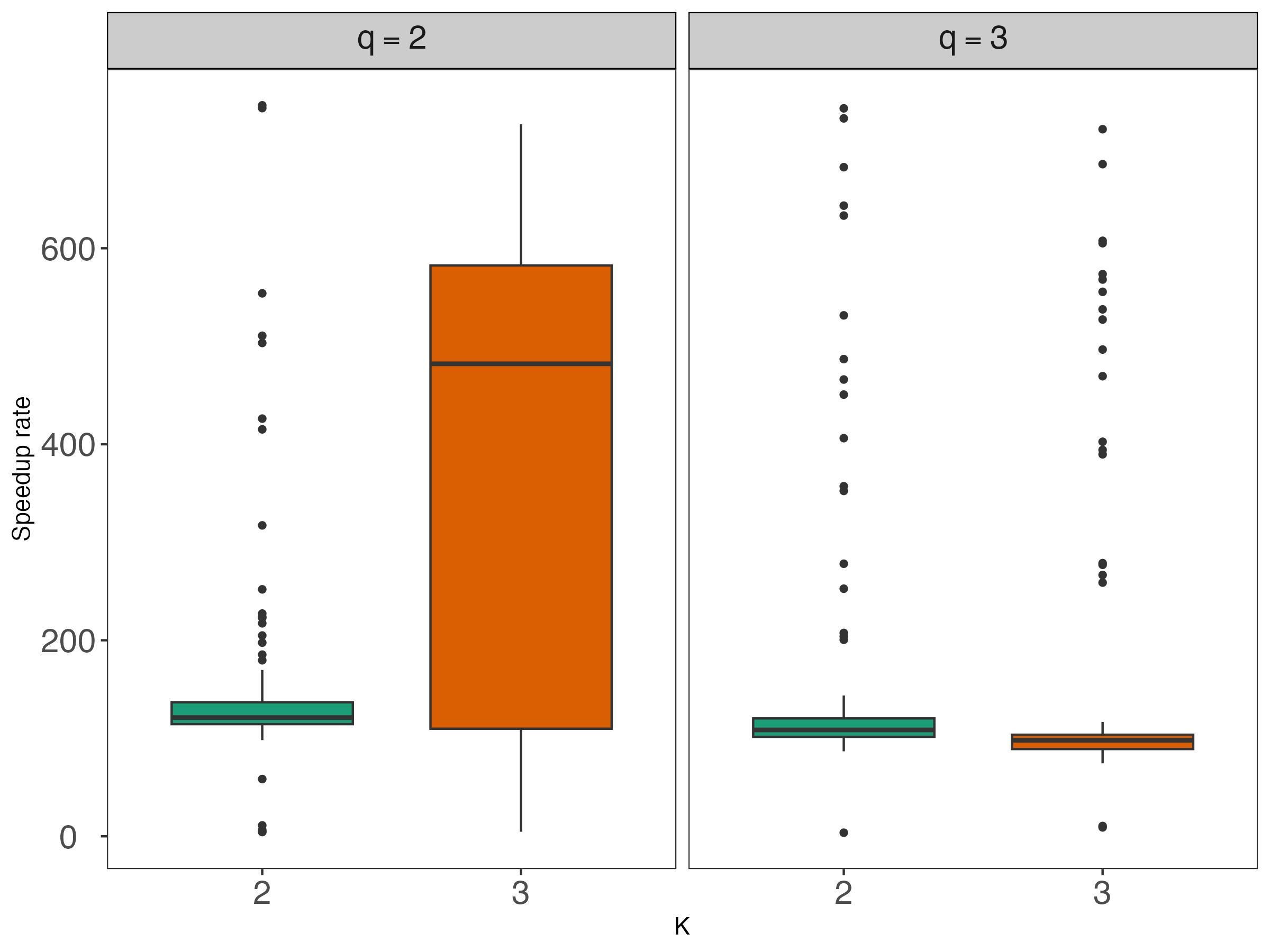}
}
}
    \caption{Boxplots of the time speedup of GMMFAD relative to EMMIX for (a) $n=300,p=10$ and (b) $n=150,p=150$, with colors indicating the true number of clusters $K=2,3$.}
    \label{time_speedup}
\end{figure}

More importantly, without the loss of the clustering and estimation accuracies, our GMMFAD significantly reduces the computational time compared to EMMIX. Figure~\ref{fig:speedup-p10} shows the relative speed of GMMFAD to EMMIX for the correct models, where we can see that GMMFAD exhibits remarkable time speedup relative to EMMIX. The computational efficiency of GMMFAD enhances with increasing $p$ as displayed in Figure \ref{fig:speedup-p150} where GMMFAD and EMMIX were fitted to simulated data with larger dimensions of $n=p=150$. Meanwhile, GMMFAD still maintained desirable estimation results and produced more accurate estimates for the loading matrix $\Lambda_k$ compared to EMMIX for this high-dimensional case, as indicated by Figure S2 
of the Supplement. In summary, our GMMFAD is able to implement the Gassian mixture of factor analyzers for large data with more efficient computations.

\section{statistical analysis of cancer data} \label{application}
\subsection{Wisconsin breast cancer data} 
The Wisconsin breast  cancer (diagnostic) dataset \citep{Street_etal1993}
(publicly available at the \href{https://archive.ics.uci.edu/dataset/17/breast+cancer+wisconsin+diagnostic}{UCI machine learning repository}) contains 569 instances with 30 features of an examination of a breast mass. The features are computed from a digitized image of a fine needle aspirate (FNA), which is a minimally invasive diagnostic procedure used to extract cellular material from a suspicious breast lump or lesion using a thin, hollow needle. The collected cells are then examined under a microscope to assess for malignancy, providing crucial insights into the presence and type of the breast cancer. The features describe the characteristics of the cell nuclei present in the image. These include the radii, area, smoothness, perimeter, compactness, texture etc of the nuclei. Figure S3 
of the Supplement displays the distributions of randomly selected features of the data which exhibit large skewness, so we normalize the dataset using a Gaussian distributional transform (GDT) \citep{zhuetal21}. The goal is to intrinsically classify the tumors into malignant or benign, hence, the number of groups is assumed to be $K=2$.

Then, we fit the transformed data using both GMMFAD and EMMIX with a common number of factors $q$ up to 25, and compared the result with the target variable. GMMFAD and EMMIX obtain a $q$ of 18 and 19, along with ARI values of 0.75 and 0.62, respectively, indicating the higher accuracy of GMMFAD for clustering this cancer data.  Like many medical tests, the reliability of applied methods for the purpose of medical diagnosis is assessed by evaluating the sensitivity and specificity \citep{Street_etal1993}. With the malignant group considered the positive class, the sensitivity and specificity of our method are respectively $0.915$ and $0.944$ as given in Table \ref{tab:metrics}, demonstrating that GMMFAD is more effective for medical diagnostic purposes. 
\begin{table}[h!]
\setlength{\tabcolsep}{2.75pt} 
\setlength\extrarowheight{3pt}
    \centering
    \caption{Performance metrics of GMMFAD and EMMIX for the breast cancer data.}
    \label{tab:metrics}
    \begin{tabular}{c|ccccc}
        \hline 
        \hline
        &\textbf{ARI} & \textbf{Accuracy} & \textbf{Sensitivity} & \textbf{Specificity}  & \textbf{Kappa} \\ \hline 
                GMMFAD & 0.750 &  0.933 & 0.915& 0.944  & 0.848 \\ \hline
                EMMIX & 0.6213 & 0.8946 & 0.9104 & 0.8852 & 0.7791 \\ \hline\hline
    \end{tabular}
\end{table}

\begin{table}[h!]
\setlength{\tabcolsep}{2pt} 
\setlength\extrarowheight{2pt}
    \centering
    \caption{Performance metrics of GMMFAD and GMMFAD-q for the breast cancer data.}
    \label{tab:metrics3}
    \begin{tabular}{c|ccccc}
        \hline 
        \hline
        &\textbf{Optimal $\bm q$} &\textbf{ARI} & \textbf{Accuracy} & \textbf{Sensitivity} & \textbf{Specificity}   \\ \hline 
                GMMFAD  & $q= 18$            & 0.75 & 0.93 & 0.92  & 0.94  \\ \hline
                GMMFAD-q & $\bm{q}= (19, 16)$ & 0.76 & 0.94 & 0.93  & 0.94  \\ \hline \hline
    \end{tabular}
\end{table}

\begin{table*}[!ht]
\setlength{\tabcolsep}{5.5pt} 
\setlength\extrarowheight{2.5pt}
\centering
\caption{Estimated factor loadings ($F_1,F_2,\ldots,F_{18}$) for the benign group identified by GMMFAD. For clarity of presentation, values in the interval $(-0.1,0.1)$ are suppressed in the table.}
\resizebox{\textwidth}{!}{ 
\begin{tabular}{c|c|c|c|c|c|c|c|c|c|c|c|c|c|c|c|c|c}
  \hline
\hline
$\bm{F_{1}}$ &
$\bm{F_{2}}$ &
$\bm{F_{3}}$ &
$\bm{F_{4}}$ &
$\bm{F_{5}}$ &
$\bm{F_{6}}$ &
$\bm{F_{7}}$ &
$\bm{F_{8}}$ &
$\bm{F_{9}}$ &
$\bm{F_{10}}$ &
$\bm{F_{11}}$ &
$\bm{F_{12}}$ &
$\bm{F_{13}}$ &
$\bm{F_{14}}$ &
$\bm{F_{15}}$ &
$\bm{F_{16}}$ &
$\bm{F_{17}}$ &
$\bm{F_{18}}$ \\
\hline
-0.92 & -0.35  &  &  &  &  &  & 0.12 &  &  &  &  &  &  & & & & \\ 
  -0.10 &      & -0.65 &  & 0.62 & & 0.10 &  & 0.16 & 0.27 &  & 0.17 & 0.10 &  &  &  &  &  \\ 
  -0.94 & -0.29 & 0.01 &  & &  &  & 0.12 &  &  &  &  &  & -0.01 &  &  &  &  \\ 
  -0.92 & -0.36 &  &  &  &  &  & 0.11 &  &  &  &  &  & &  & & &  \\ 
        & 0.50 & 0.13 & -0.32 & -0.15 & -0.53 & -0.32 & 0.21 & & &  &  & 0.20 & & &  & -0.24 & \\ 
  -0.43 & 0.79 &  &  &  & -0.15 &  &  & 0.16 & -0.13 & -0.11 & 0.16 & 0.14 &  & &  & & \\ 
  -0.59 & 0.72 &  & 0.15 &  & 0.14 & -0.24 &  &  &  &  &  &  &  &  &  &  &  \\ 
  -0.70 & 0.49 &  &  & -0.12 & & -0.29 & 0.16 &  & & 0.20 & 0.19 & &  &  &  &  &  \\ 
        & 0.47 & & -0.76 & -0.12 & 0.25 &  & 0.24 & 0.22 &  & & &  &  & & &  & \\ 
  0.29  & 0.82 &  &  & -0.12 & -0.20 & 0.15 & 0.14 & & -0.10 & 0.10 &  & 0.30 &  &  & & & \\ 
  -0.13 & 0.13 & -0.69 & -0.34 & -0.45 & -0.18 &  & -0.22 &  &  &  &  &  & 0.22 &  &  &  &  \\ 
  0.24  & 0.15 & -0.84 & -0.10 & 0.26 &  & -0.11 & 0.25 & -0.12 & -0.20 &  &  &  &  & & &  &  \\ 
  -0.23 & 0.28 & -0.70 & -0.26 & -0.44 & -0.11 &  & -0.29 &  & & &  & & -0.11 &  & & & \\ 
  -0.42 &  & -0.66 & -0.30 & -0.42 & -0.16 &  & -0.16 &  &  &  &  &  & 0.18 & &  &  &  \\ 
  0.39 & 0.45 & -0.27 &  & -0.16 & -0.33 & -0.26 & 0.13 &  & 0.26 & -0.16 & 0.15 & -0.11 &  &  & 0.11 & 0.23 & 0.12 \\ 
  -0.31 & 0.81 & -0.25 & 0.15 & -0.11 & 0.11 &  &  &  &  & -0.25 & 0.14 &  &  &  &  &  &  \\ 
  -0.41 & 0.76 & -0.16 & 0.26 &  & 0.28 & -0.16 &  &  & 0.12 &  &  &  &  & 0.13 &  &  &  \\ 
  -0.41 & 0.64 & -0.27 & 0.11 & -0.25 & 0.11 & -0.28 & 0.15 &  &  & 0.16 & 0.23 &  &  & 0.16 &  &  &  \\ 
  0.28 & 0.30 & -0.38 & -0.37 & -0.25 & 0.18 &  &  & -0.40 &  & -0.10 & 0.35 & 0.21 &  & & 0.15 & & \\ 
       & 0.79 & -0.29 & 0.19 & -0.22 &  & 0.35 & 0.21 &  & 0.12 &  &  &  &  &  &  &  &  \\ 
  -0.93 & -0.33 & &  &  & &  &  & &  & &  &  & & &  & &  \\ 
        &  & -0.57 &  & 0.79 &  & &  &  &  &  &  & & &  &  &  & \\ 
  -0.96 & -0.23 &  &  &  &  &  &  &  &  &  &  &  &  &  &  &  &  \\ 
  -0.93 & -0.34 &  &  &  &  & &  &  &  &  &  &  &  &  &  &  &  \\ 
        & 0.55 & 0.35 & -0.25 & 0.13 & -0.61 &  &  &  &  &  &  &  &  & &  &  &  \\ 
  -0.50 & 0.76 & 0.17 &  & 0.13 & & 0.12 & -0.14 & 0.15 & -0.15 & -0.16 & &  & &  & & & \\ 
  -0.58 & 0.70 & 0.13 & 0.18 & 0.16 & 0.19 & -0.13 & -0.11 & &  & & -0.12 & & & & & & \\ 
  -0.72 & 0.52 & 0.18 &  &  &  & -0.20 &  &  & -0.11 & 0.27 & 0.12 & -0.13 & & &  &  &  \\ 
        & 0.39 & 0.39 & -0.71 & 0.21 & 0.20 & 0.15 & -0.13 & -0.24 &  &  & &  & &  &  &  &  \\ 
        & 0.85 & 0.19 & 0.13 & & -0.21 & 0.39 &  &  &  & 0.10 & -0.11 &  &  & &  &  &  \\ 
   \hline
   \hline
\end{tabular}
}
\label{factor loadings benign}
\end{table*}
To further characterize the identified clusters, Table \ref{factor loadings benign} presents the fitted factor loadings of the estimated benign group from GMMFAD, with values that are not negligible (outside the interval $(-0.1,0.1)$). The 18 factors for the benign group explain over $98\%$ of the total data variability within this cluster, where the first few factors are viewed as contrasts, with the first  factor exhibiting mostly substantial to very high negative influence on the features, while the second factor exhibits mostly substantial positive influence on some of the features. Distinguished trend can be seen from Table \ref{factor loadings malignant} with fitted factor loadings for the malignant group where the factors together explain about $90\%$ of the total variation. The first factor also contributes negatively to the observation, exacting very strong influence on half of the features, but the significant contributions come from a different set of features compared to the first loadings of the benign group. The second factor shows contrast across the features with the strongest influences being positive. The third and fourth factor loadings of the malignant group have correlations with less features compared to the benign cluster. In summary, the factor loadings for the two breast cancer groups are collectively distinct, providing additional characterization to the dispersion of the tumor samples for different breast cancer types.

\begin{table*}[!ht]
\setlength{\tabcolsep}{5.5pt} 
\setlength\extrarowheight{2.5pt}
\centering
\caption{Estimated factor loadings ($F_1,F_2,\ldots,F_{18}$) for the malignant group identified by GMMFAD. For clarity of presentation, values in the interval $(-0.1,0.1)$ are suppressed in the table. }
 
\begin{tabular}{c|c|c|c|c|c|c|c|c|c|c|c|c|c|c|c|c|c}
  \hline
\hline
$\bm{F_{1}}$ &
$\bm{F_{2}}$ &
$\bm{F_{3}}$ &
$\bm{F_{4}}$ &
$\bm{F_{5}}$ &
$\bm{F_{6}}$ &
$\bm{F_{7}}$ &
$\bm{F_{8}}$ &
$\bm{F_{9}}$ &
$\bm{F_{10}}$ &
$\bm{F_{11}}$ &
$\bm{F_{12}}$ &
$\bm{F_{13}}$ &
$\bm{F_{14}}$ &
$\bm{F_{15}}$ &
$\bm{F_{16}}$ &
$\bm{F_{17}}$ &
$\bm{F_{18}}$ \\
\hline
-0.89 & -0.38 & & & & & -0.10 & -0.10 & & & & & & & & & & \\ 
-0.33 &  &  & 0.90 & &  &  & 0.16 & -0.20 &  &  &  & &  & & &  & \\ 
-0.91 & -0.31 & & &  & & -0.10 & -0.10 & -0.11 & & & &  & &  & & \quad\quad\quad &  \\ 
-0.89 & -0.39 & &  & & &  & & & & &  & & & & &  & \\ 
-0.17 & 0.77 & & -0.19 & 0.28 & -0.38 & -0.23 & & -0.17 & &  &  & 0.11 &  &  & &  & \\ 
-0.54 & 0.76 & 0.18 &  &  &  &  & &  &  &  & -0.19 &  &  &  &  & & \\ 
-0.79 & 0.47 & 0.12 &  & 0.12 &  & -0.20 &  &  & -0.14 &  &  & & & & 0.13 &  & \\ 
-0.88 & 0.26 & & -0.13 & 0.11 & & -0.21 & -0.10 & & & -0.10 & & & & & 0.15 & &  \\ 
-0.23 & 0.72 & &  & -0.30 & -0.23 &  & -0.10 & -0.11 &  & &  & & 0.31 &  & 0.12 & & -0.14 \\ 
      & 0.89 & 0.12 &  & 0.23 & -0.15 & 0.15 & -0.14 &  & &  &  & -0.22 &  & &  &  & \\ 
-0.82 & & -0.43 & -0.11 &  & -0.20 & 0.24 &  &  & -0.11 &  &  &  & &  & & &  \\ 
-0.18 & 0.29 & -0.68 & 0.45 &  & -0.12 & -0.16 & -0.33 & 0.22 &  & &  &  &  &  &  &  & \\ 
-0.83 &  & -0.44 &  &  & -0.14 & 0.21 & & & -0.12 &  & &  & &  &  &  &  \\ 
-0.89 & -0.17 & -0.28 &  &  & -0.14 & 0.19 &  & &  & &  & &  & &  & & \\ 
-0.19 & 0.49 & -0.69 &  & 0.16 &  & -0.18 & 0.35 &  & 0.21 &  &  &  &  &  & &  &  \\ 
-0.49 & 0.74 &  & &  & 0.33 & 0.14 & & & & & -0.19 &  & & 0.12 & & & \\ 
-0.61 & 0.59 & -0.20 &  & & 0.38 & -0.11 &  & & -0.22 &  &  &  & &  & & & \\ 
-0.59 & 0.37 & -0.41 & &  & 0.25 &  & 0.12 & 0.15 & & -0.28 & & & 0.14 & 0.21 &  & & \\ 
-0.27 & 0.54 & -0.29 & -0.13 & -0.70 &  &  &  &  &  &  &  &  & -0.13 &  &  &  & \\ 
-0.31 & 0.81 & -0.20 &  & & 0.18 & 0.28 & -0.12 & -0.12 & 0.16 & &  & 0.12 & &  & & & \\ 
-0.91 & -0.33 & 0.18 &  &  &  &  &  &  &  &  & &  &  &  &  &  &  \\ 
-0.14 & 0.14 & 0.16 & 0.92 & & -0.19 & & & & & & & & &  &  &  & \\ 
-0.93 & -0.26 & 0.21 & & & & & & & &  &  & & & & &  & \\ 
-0.90 & -0.35 & 0.17 & & & & & & & & & & & & &  &  & \\ 
      & 0.69 & 0.19 & & 0.31 & -0.32 & -0.22 & 0.33 & & 0.15 & 0.19 & & &  &  &  & &\\
-0.30 & 0.73 & 0.50 & 0.11 & & 0.15 & 0.13 & & 0.14 & & & -0.18 & &  &  & &  & \\
-0.50 & 0.60 & 0.44 & 0.10 & & 0.20 & -0.12 & & & -0.21 & 0.17 & 0.12 & &  & -0.10 & & &\\ 
-0.71 & 0.39 & 0.41 &  & &  & -0.18 & & 0.21 & & -0.28 & & &  & & &  & \\ 
      & 0.58 & 0.50 & & -0.56 & -0.23 &  & & &  & & & & 0.13 &  & & & \\ 
      & 0.80 & 0.44 & & 0.16 & & 0.27 & & & 0.17 & & 0.12 & &  &  &  &  & \\ 
\hline
\hline
\end{tabular}
\label{factor loadings malignant}
\end{table*}
We further fit the data with GMMFAD-q without the assumption of the same number of factors across the groups. From Table \ref{tab:metrics3}, the best model selected by BIC in this case is $\bm{q}_{opt}= (19, 16)$, with an ARI of $0.76$, and the values of accuracy, sensitivity, and specificity are $0.94, 0.93$ and $0.94 $, respectively, mostly higher than the corresponding values when $q$ is fixed. Our approach with varied number of factors therefore exhibits an added flexibility in probing more complex latent structures among mixed data and thus stronger potential to increase accuracy in clustering and describing complex datasets.  

\subsection{Lymphoma gene expression data}
Lymphoma is a group of  lymph cancers that affect the lymphatic system. The lymphoma dataset we consider is available in the R package \texttt{spls} \citep{chung2019spls, Chungetal2010}, which comprises gene expression profiles for $n=62$ patients, categorized into $42$ cases of diffuse large B-cell lymphoma (DLBCL), $9$ cases of follicular lymphoma (FL), and $11$ cases of chronic lymphocytic leukemia (CLL), across $p= 4026$ genes. The class labels for DLBCL, FL, and CLL are encoded as 0, 1, and 2, respectively, in the response vector, while the predictor matrix contains the gene expression measurements. The data preprocessing requires normalization, imputation, log-transformation, and standardization to zero mean and unit variance across genes, following the methodologies outlined in \citet{dettlingetal2002,dettling2004}. For this massive dataset, our goal is to efficiently distinguish the lymphoma subtypes and summarize the variability within each of the three classes.

We fit the data using GMMFAD-q for a generalized MFA with the number of groups assumed known to be $K=3$ and the maximum number of factors of $18$. (EMMIX is impractical here given its extremely slow convergence due to the high dimension of the data ($p=4026$).) The optimal model has $\bm{q}_{opt}= (10,9,8)$ for the three estimated clusters of DLBCL, FL and CLL, respectively,  with an ARI of $0.95$, where only one point was misclassified.  Figure S4 
of the Supplement depicts the distinguished loading patterns for different disease subtypes. Figure S5 
of the Supplement shows the distribution curves of the estimated factor loadings within each cluster and we can see that the variational artifacts in these curves across the subtypes of the disease highlight the inherent distinction exhibited among the subtypes of lymphoma in lower dimensional spaces. Our method demonstrates a strong capacity to model high dimensional data especially in situation with an extremely large $p$ and $n\ll p$.

\section{Conclusion} \label{conclusion}
We propose a hybrid approach for estimating the parameters from the mixture of factor analyzers, which combines matrix-free computations with the EM algorithm. The matrix-free component significantly improves the computational efficiency of the method, particularly for high-dimensional data, while maintaining high clustering and estimation accuracy. Through simulations, the proposed method exhibits stronger clustering performance compared to the existing algorithm. Additionally, we extend the approach to a generalized model with varying numbers of factors across clusters, and apply the methods to cluster and characterize the Wisconsin breast cancer dataset and the lymphoma dataset, successfully identifying the subtypes with remarkable accuracy rates. The developed methods and algorithms pave the way to clustering data with non-Gaussian distributions, and data with more complex structures such as partial records, mixed features and measurement errors.

\section*{Acknowledgments}
The research of the first author was supported in part by the Michigan Technological University Doctoral Finishing Fellowship Award. The authors would like to appreciate the Graduate school for providing this support.

\section*{Supplementary Materials}
The following supplementary materials are available and contain:
\begin{enumerate}
\item A Supplement file 
(\texttt{figure-for-submission.pdf}) containing supplementary figures for Section \ref{methodology} and Section \ref{application} in this article.
  \item A compressed file (\texttt{code-for-submission.zip}) containing the code required to produce the simulation and data application results in this article.
\end{enumerate}

\section*{Conflict of Interest}
None.

\section*{Data availability statement}
The data used in this article are all publicly available. The Wisconsin breast cancer dataset is available at the UCI machine learning repository (\url{https://archive.ics.uci.edu/dataset/17/breast+cancer+wisconsin+diagnostic}).
The lymphoma dataset is available in the R package \texttt{spls} under the name \textit{lymphoma}.

\bibliography{ref}

\begin{thebibliography}{10}
\providecommand{\url}[1]{#1}
\csname url@rmstyle\endcsname
\providecommand{\newblock}{\relax}
\providecommand{\bibinfo}[2]{#2}
\providecommand\BIBentrySTDinterwordspacing{\spaceskip=0pt\relax}
\providecommand\BIBentryALTinterwordstretchfactor{4}
\providecommand\BIBentryALTinterwordspacing{\spaceskip=\fontdimen2\font plus
\BIBentryALTinterwordstretchfactor\fontdimen3\font minus \fontdimen4\font\relax}
\providecommand\BIBforeignlanguage[2]{{%
\expandafter\ifx\csname l@#1\endcsname\relax
\typeout{** WARNING: mytran.bst: No hyphenation pattern has been}%
\typeout{** loaded for the language `#1'. Using the pattern for}%
\typeout{** the default language instead.}%
\else
\language=\csname l@#1\endcsname
\fi
#2}}

\bibitem{otuetal2005}
H.~Otu, S.~Kolia, J.~Jones, O.~Osman, and T.~Libermann, ``Significance analysis of clustering high throughput biological data,'' 06 2005, pp. 6 pp. -- 6.

\bibitem{choietal2020}
I.~Choi, H.~Lim, H.~Cho, Y.~Oh, B.-K. Chou, H.~Bai, L.~Cheng, Y.~J. Kim, S.~Hyun, H.~Kim, J.~Shin, and G.~Lee, ``Transcriptional landscape of myogenesis from human pluripotent stem cells reveals a key role of twist1 in maintenance of skeletal muscle progenitors,'' \emph{eLife}, vol.~9, 02 2020.

\bibitem{eisenetal1998}
\BIBentryALTinterwordspacing
M.~B. Eisen, P.~T. Spellman, P.~O. Brown, and D.~Botstein, ``Cluster analysis and display of genome-wide expression patterns,'' \emph{Proceedings of the National Academy of Sciences of the United States of America}, vol.~95, no.~25, pp. 14\,863--14\,868, 1998. [Online]. Available: \url{https://doi.org/10.1073/pnas.95.25.14863}
\BIBentrySTDinterwordspacing

\bibitem{Selinskietal2008}
\BIBentryALTinterwordspacing
S.~Selinski and K.~Ickstadt, ``Cluster analysis of genetic and epidemiological data in molecular epidemiology,'' \emph{Journal of Toxicology and Environmental Health, Part A}, vol.~71, no. 11-12, pp. 835--844, 2008. [Online]. Available: \url{https://doi.org/10.1080/15287390801985828}
\BIBentrySTDinterwordspacing

\bibitem{everitt1974}
B.~Everitt, ``Cluster analysis,'' \emph{Wiley Series in Probability and Statistics}, 1974.

\bibitem{macqueen1967}
J.~MacQueen, ``Some methods for classification and analysis of multivariate observations,'' in \emph{Proceedings of the fifth Berkeley symposium on mathematical statistics and probability}, vol.~1.\hskip 1em plus 0.5em minus 0.4em\relax Oakland, CA, USA, 1967, pp. 281--297.

\bibitem{bezdek1981}
J.~C. Bezdek, ``Pattern recognition with fuzzy objective function algorithms,'' \emph{Plenum Press}, 1981.

\bibitem{comaniciuetal2002}
D.~Comaniciu and P.~Meer, ``Mean shift: A robust approach toward feature space analysis,'' \emph{IEEE Transactions on Pattern Analysis and Machine Intelligence}, vol.~24, no.~5, pp. 603--619, 2002.

\bibitem{ngetal2002}
\BIBentryALTinterwordspacing
A.~Ng, M.~Jordan, and Y.~Weiss, ``On spectral clustering: Analysis and an algorithm,'' in \emph{Advances in Neural Information Processing Systems}, T.~Dietterich, S.~Becker, and Z.~Ghahramani, Eds., vol.~14.\hskip 1em plus 0.5em minus 0.4em\relax MIT Press, 2001. [Online]. Available: \url{https://proceedings.neurips.cc/paper_files/paper/2001/file/801272ee79cfde7fa5960571fee36b9b-Paper.pdf}
\BIBentrySTDinterwordspacing

\bibitem{mclachlanpeel2000}
G.~J. McLachlan and D.~Peel, \emph{Finite Mixture Models}.\hskip 1em plus 0.5em minus 0.4em\relax New York: John Wiley \& Sons, 2000.

\bibitem{mclachlanetal2003}
\BIBentryALTinterwordspacing
G.~J. McLachlan, D.~Peel, and R.~W. Bean, ``Modelling high-dimensional data by mixtures of factor analyzers,'' \emph{Computational Statistics \& Data Analysis}, vol.~41, no. 3-4, pp. 379--388, 2003. [Online]. Available: \url{https://EconPapers.repec.org/RePEc:eee:csdana:v:41:y:2003:i:3-4:p:379-388}
\BIBentrySTDinterwordspacing

\bibitem{melnykovetal2012}
\BIBentryALTinterwordspacing
V.~Melnykov, W.-C. Chen, and R.~Maitra, ``Mixsim: An r package for simulating data to study performance of clustering algorithms,'' \emph{Journal of Statistical Software}, vol.~51, no.~12, p. 1–25, 2012. [Online]. Available: \url{https://www.jstatsoft.org/index.php/jss/article/view/v051i12}
\BIBentrySTDinterwordspacing

\bibitem{dempsteretal1977}
A.~P. Dempster, N.~M. Laird, and D.~B. Rubin, ``Maximum likelihood from incomplete data via the em algorithm,'' \emph{Journal of the Royal Statistical Society: Series B (Methodological)}, vol.~39, no.~1, pp. 1--22, 1977.

\bibitem{tippingBishop1999}
M.~E. Tipping and C.~M. Bishop, ``Mixtures of probabilistic principal component analysers,'' \emph{Neural Computation}, vol.~11, no.~2, pp. 443--482, 1999.

\bibitem{ghahramaniHinton1996}
Z.~Ghahramani and G.~E. Hinton, ``The em algorithm for mixtures of factor analyzers,'' \emph{Technical Report CRG-TR-96-1}, 1996.

\bibitem{andrewsetal2011}
\BIBentryALTinterwordspacing
J.~L. Andrews and P.~D. McNicholas, ``Mixtures of modified t-factor analyzers for model-based clustering, classification, and discriminant analysis,'' \emph{Journal of Statistical Planning and Inference}, vol. 141, no.~4, pp. 1479--1486, 2011. [Online]. Available: \url{https://www.sciencedirect.com/science/article/pii/S0378375810004830}
\BIBentrySTDinterwordspacing

\bibitem{mcnicholas2008}
\BIBentryALTinterwordspacing
P.~McNicholas and T.~Murphy, ``Parsimonious gaussian mixture models,'' \emph{Stat Comput}, vol.~18, p. 285–296, 2008. [Online]. Available: \url{https://doi.org/10.1007/s11222-008-9056-0}
\BIBentrySTDinterwordspacing

\bibitem{daietal2020}
F.~Dai, S.~Dutta, and R.~Maitra, ``A matrix-free likelihood method for exploratory factor analysis of high-dimensional gaussian data,'' \emph{J Comput Graph Stat}, vol.~29, no.~3, pp. 675--680, 2020.

\bibitem{redneretal1984}
R.~A. Redner and H.~F. Walker, ``Mixture densities, maximum likelihood and the em algorithm,'' \emph{SIAM Review}, vol.~26, no.~2, pp. 195--239, 1984.

\bibitem{bishop2006}
C.~M. Bishop, \emph{Pattern Recognition and Machine Learning}.\hskip 1em plus 0.5em minus 0.4em\relax New York: Springer, 2006.

\bibitem{baglamareichel2005}
\BIBentryALTinterwordspacing
J.~Baglama and L.~Reichel, ``Augmented implicitly restarted lanczos bidiagonalization methods,'' \emph{SIAM Journal on Scientific Computing}, vol.~27, no.~1, pp. 19--42, 2005. [Online]. Available: \url{https://doi.org/10.1137/04060593X}
\BIBentrySTDinterwordspacing

\bibitem{byrd1995}
R.~H. Byrd, J.~N. P.~Lu, and C.~Zhu, ``A limited memory algorithm for bound constrained optimization,'' \emph{SIAM Journal on Scientific Computing}, vol.~16, pp. 1190--1208, 1995.

\bibitem{Biernackietal2003}
\BIBentryALTinterwordspacing
G.~G. Christophe~Biernacki, Gilles~Celeux, ``Choosing starting values for the em algorithm for getting the highest likelihood in multivariate gaussian mixture models,'' \emph{Computational Statistics \& Data Analysis}, vol.~41, pp. 561--575, 2003. [Online]. Available: \url{https://doi.org/10.1016/S0167-9473(02)00163-9}
\BIBentrySTDinterwordspacing

\bibitem{Maitra2013}
R.~Maitra, ``On the expectation-maximization algorithm for rice-rayleigh mixtures with application to noise parameter estimation in magnitude mr datasets,'' \emph{Sankhy\=a: The Indian Journal of Statistics, Series B}, vol.~75, no.~2, pp. 293--318, 2013.

\bibitem{schwarz1978}
\BIBentryALTinterwordspacing
G.~Schwarz, ``Estimating the dimension of a model,'' \emph{The Annals of Statistics}, vol.~6, no.~2, pp. 461--464, 1978. [Online]. Available: \url{http://www.jstor.org/stable/2958889}
\BIBentrySTDinterwordspacing

\bibitem{lawleymaxwell1971}
D.~N. Lawley and A.~E. Maxwell, \emph{Factor Analysis as a Statistical Method}, 2nd~ed.\hskip 1em plus 0.5em minus 0.4em\relax London: Butterworths, 1971.

\bibitem{maitraandmelnykov2010}
\BIBentryALTinterwordspacing
R.~Maitra and V.~Melnykov, ``Simulating data to study performance of finite mixture modeling and clustering algorithms,'' \emph{Journal of Computational and Graphical Statistics}, vol.~19, no.~2, pp. 354--376, 2010. [Online]. Available: \url{https://doi.org/10.1198/jcgs.2009.08054}
\BIBentrySTDinterwordspacing

\bibitem{Zhuetal2022}
\BIBentryALTinterwordspacing
F.~D. Yifan~Zhu and R.~Maitra, ``Fully three-dimensional radial visualization,'' \emph{Journal of Computational and Graphical Statistics}, vol.~31, no.~3, pp. 935--944, 2022. [Online]. Available: \url{https://doi.org/10.1080/10618600.2021.2020129}
\BIBentrySTDinterwordspacing

\bibitem{R}
\BIBentryALTinterwordspacing
{R Core Team}, \emph{R: A Language and Environment for Statistical Computing}, R Foundation for Statistical Computing, Vienna, Austria, 2019. [Online]. Available: \url{https://www.R-project.org/}
\BIBentrySTDinterwordspacing

\bibitem{Street_etal1993}
\BIBentryALTinterwordspacing
W.~N. Street, W.~H. Wolberg, and O.~L. Mangasarian, ``Nuclear feature extraction for breast tumor diagnosis,'' in \emph{Electronic imaging}, 1993. [Online]. Available: \url{https://api.semanticscholar.org/CorpusID:14922543}
\BIBentrySTDinterwordspacing

\bibitem{zhuetal21}
\BIBentryALTinterwordspacing
Y.~Zhu, F.~Dai, and R.~Maitra, ``Visualization of labeled mixed-featured datasets,'' 2021. [Online]. Available: \url{https://arxiv.org/abs/1904.06366}
\BIBentrySTDinterwordspacing

\bibitem{chung2019spls}
D.~Chung, H.~Chun, and S.~Keles, \emph{spls: Sparse Partial Least Squares (spls) Regression and Classification}, Comprehensive R Archive Network (CRAN), https://CRAN.R-project.org/package=spls, 2019, r package version 2.2-3.

\bibitem{Chungetal2010}
\BIBentryALTinterwordspacing
D.~Chung and S.~Keles, ``Sparse partial least squares classification for high dimensional data,'' \emph{Statistical Applications in Genetics and Molecular Biology}, vol.~9, no.~1, 2010. [Online]. Available: \url{https://doi.org/10.2202/1544-6115.1492}
\BIBentrySTDinterwordspacing

\bibitem{dettlingetal2002}
M.~Dettling and P.~Bühlmann, ``Supervised clustering of genes,'' \emph{Genome Biology}, vol.~3, no.~12, pp. research0069--1, 2002.

\bibitem{dettling2004}
\BIBentryALTinterwordspacing
M.~Dettling, ``Bagboosting for tumor classification with gene expression data,'' \emph{Bioinformatics}, vol.~20, no.~18, pp. 3583--3593, 10 2004. [Online]. Available: \url{https://doi.org/10.1093/bioinformatics/bth447}
\BIBentrySTDinterwordspacing

\end{thebibliography}
\bibliographystyle{mytran}
\clearpage
\onecolumn
\begin{center}
\textbf{\large Supplement to \\
``A Hybrid Mixture Approach for Clustering and Characterizing Cancer Data" 
}
\end{center}
\setcounter{equation}{0}
\setcounter{figure}{0}
\setcounter{table}{0}
\setcounter{page}{1}
\setcounter{section}{0}
\makeatletter
\renewcommand{\thesection}{S\arabic{section}}
\renewcommand{\thesubsection}{\thesection.\arabic{subsection}}
\renewcommand{\theequation}{S\arabic{equation}}
\renewcommand{\thefigure}{S\arabic{figure}}
\renewcommand{\thetable}{S\arabic{table}}

\section{Supplementary Figures for Methods and Algorithms}
\label{sec:supp-meth}
\begin{figure}[h!]
  \centering
\mbox{
\subfloat[$\hat{\Lambda}_k\hat{\Lambda}_k^\top, K=2, q = 2,3$]{
\includegraphics[width=0.32\linewidth]{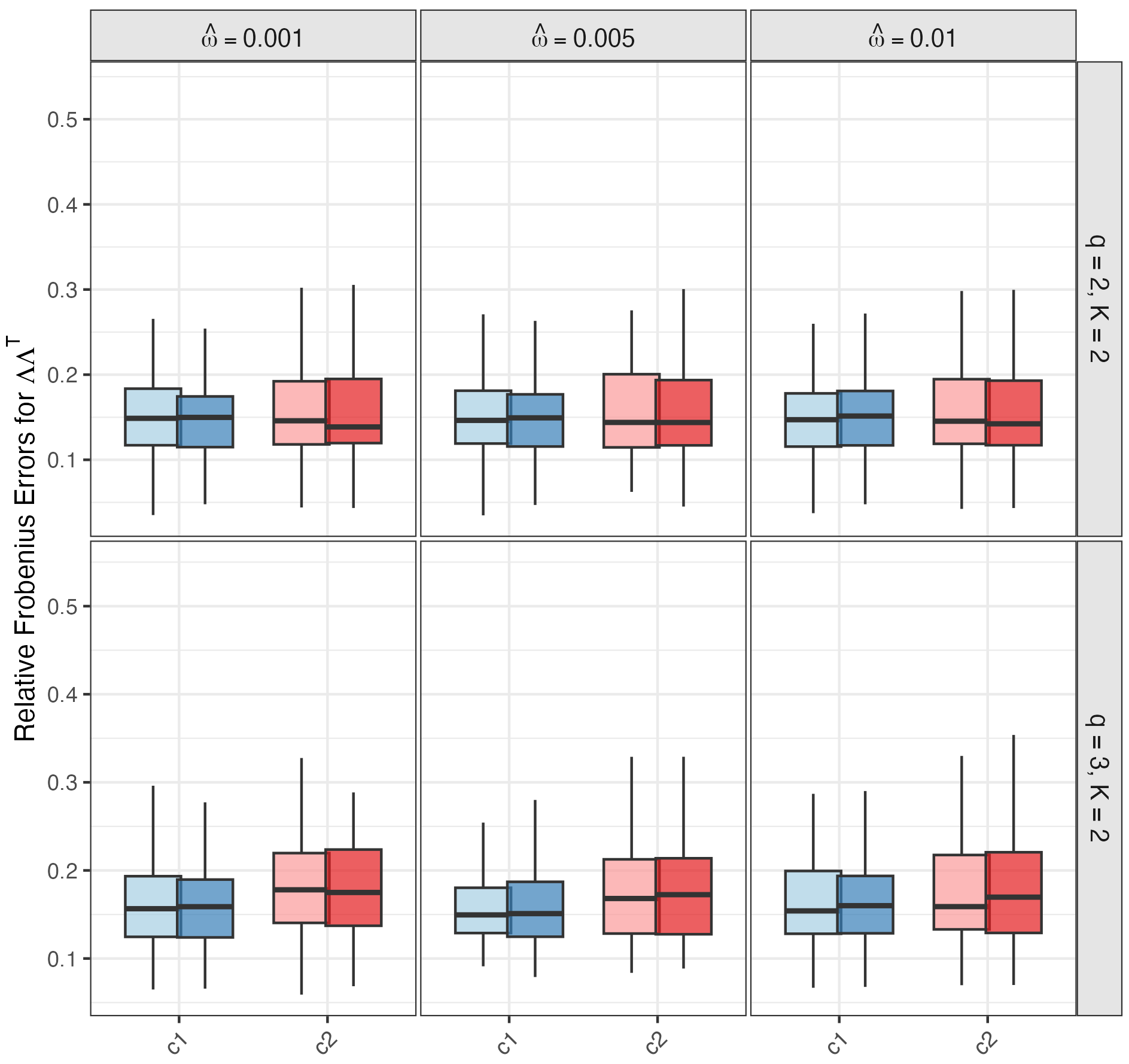}
}%
\subfloat[$\hat{\Psi}_k, K=2,q = 2,3$]{
\includegraphics[width=0.32\linewidth]{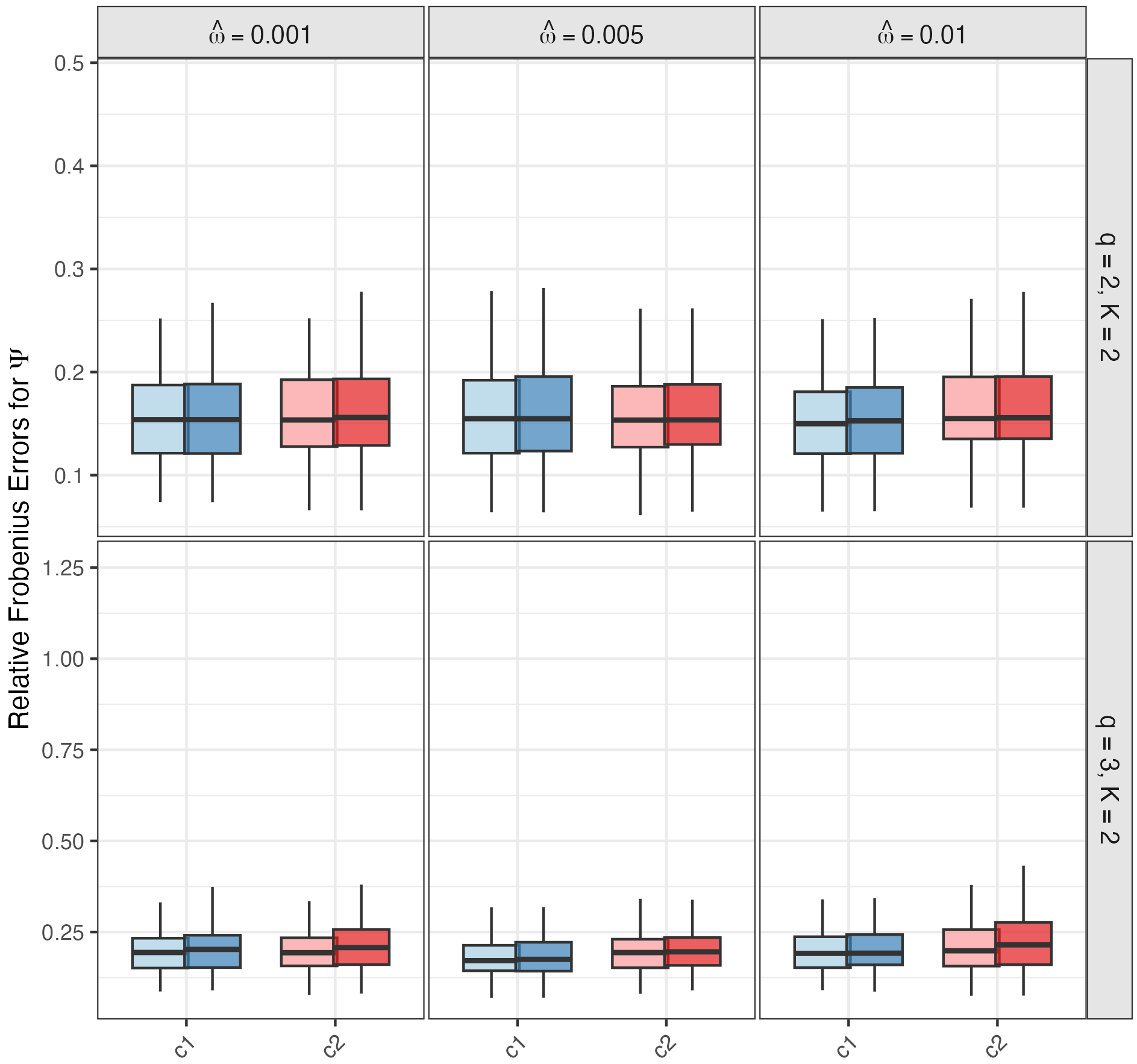}
}%
\subfloat[$\hat{\bmu}_k, K=2,q = 2,3$]{
\includegraphics[width=0.32\linewidth]{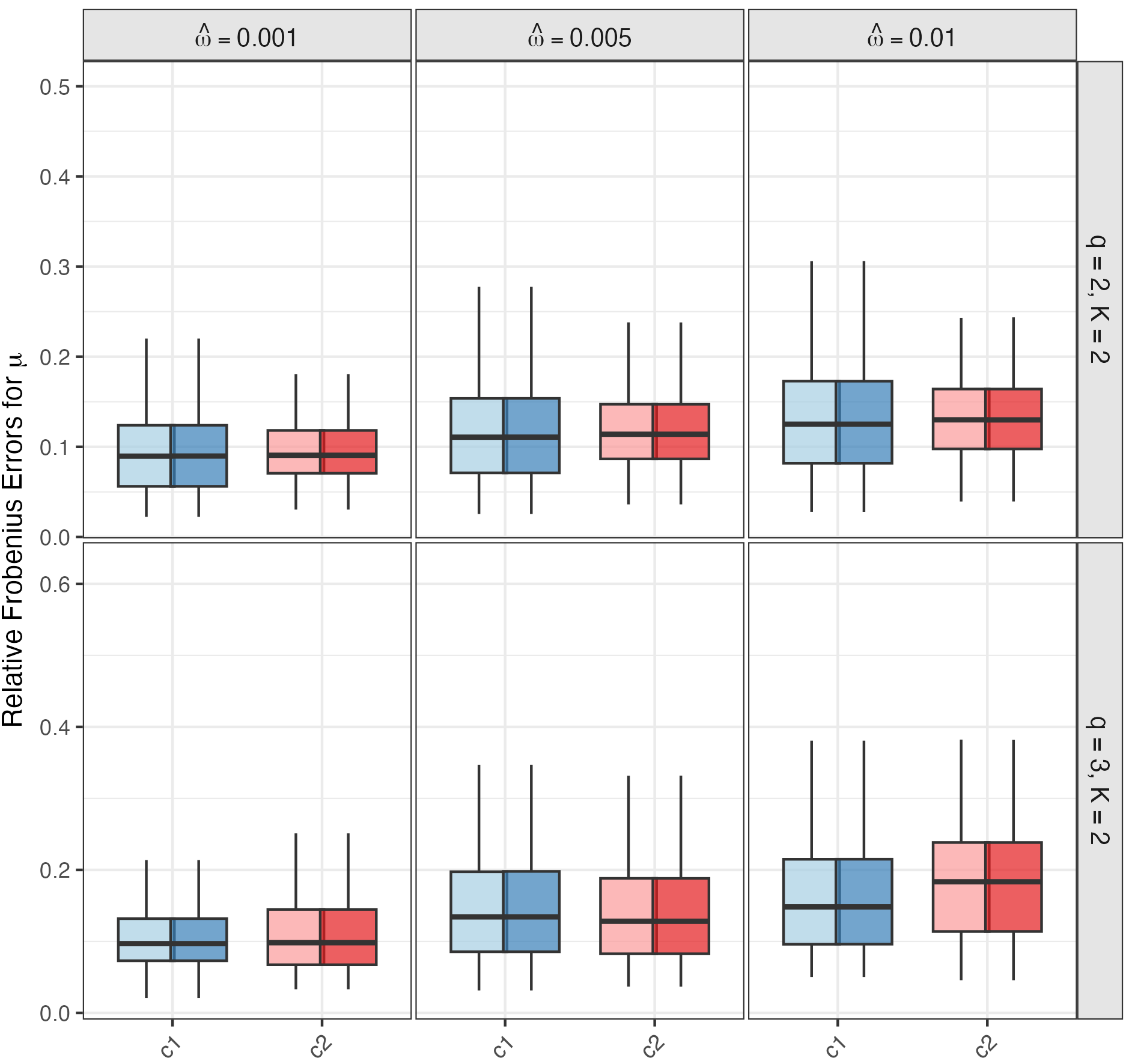}
}
}\\
\mbox{
\subfloat[$\hat{\Lambda}_k\hat{\Lambda}_k^\top, K=3,q = 2,3$]{
\includegraphics[width=0.32\linewidth]{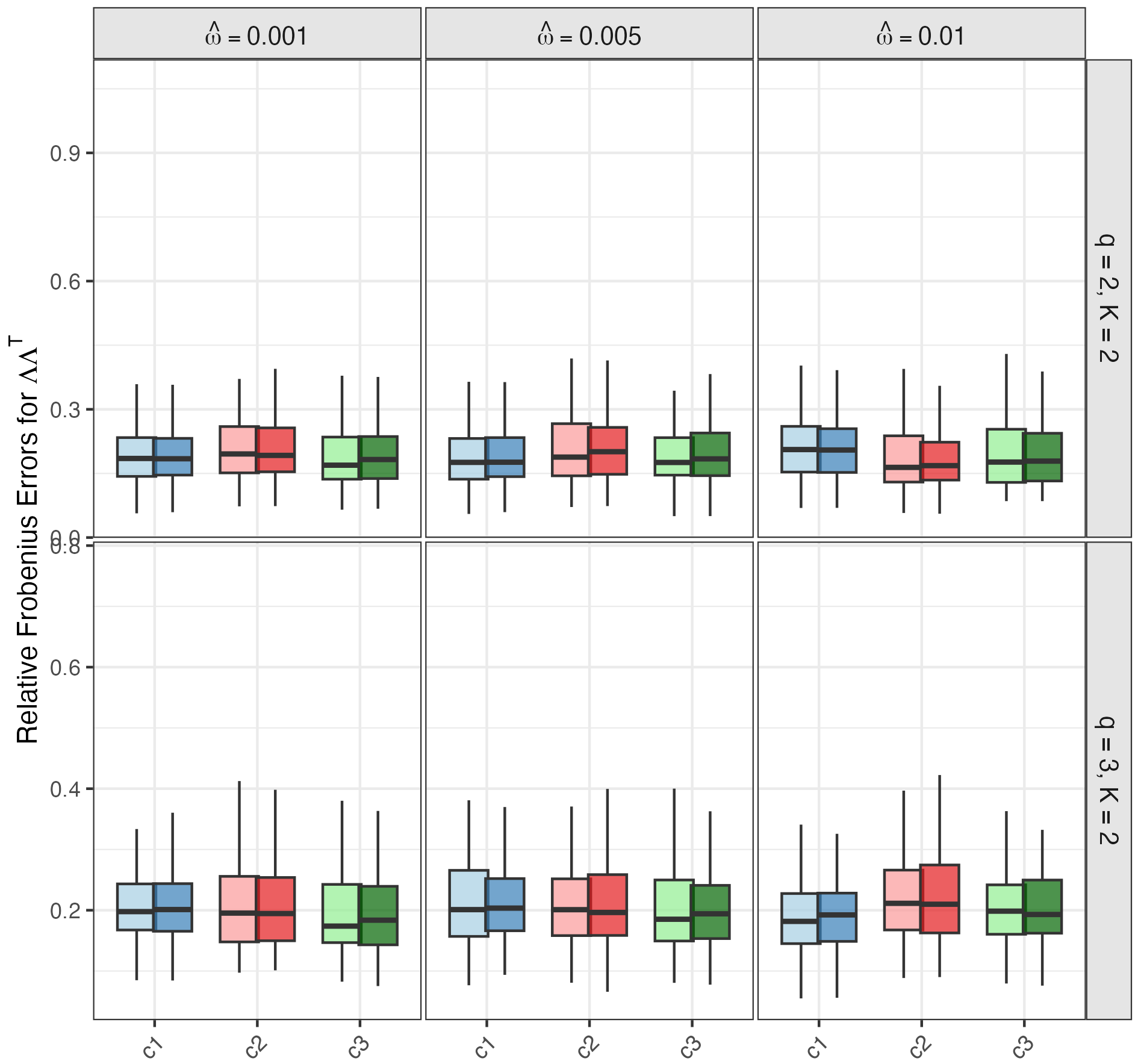}
}%
\subfloat[$\hat{\Psi}_k, K=3,q = 2,3$]{
\includegraphics[width=0.32\linewidth]{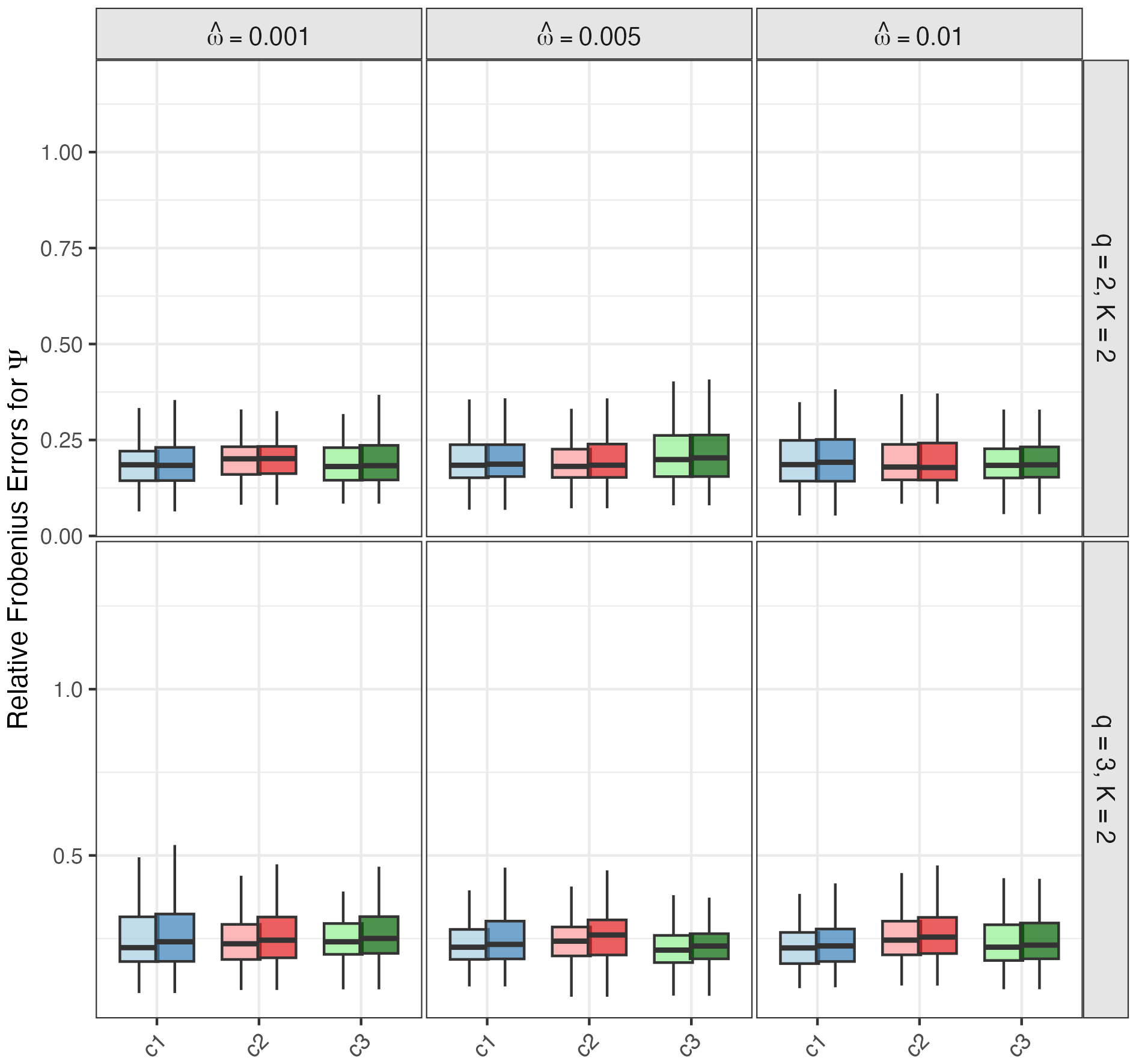}
}%
\subfloat[$\hat{\bmu}_k, K=3,q = 2,3$]{
\includegraphics[width=0.32\linewidth]{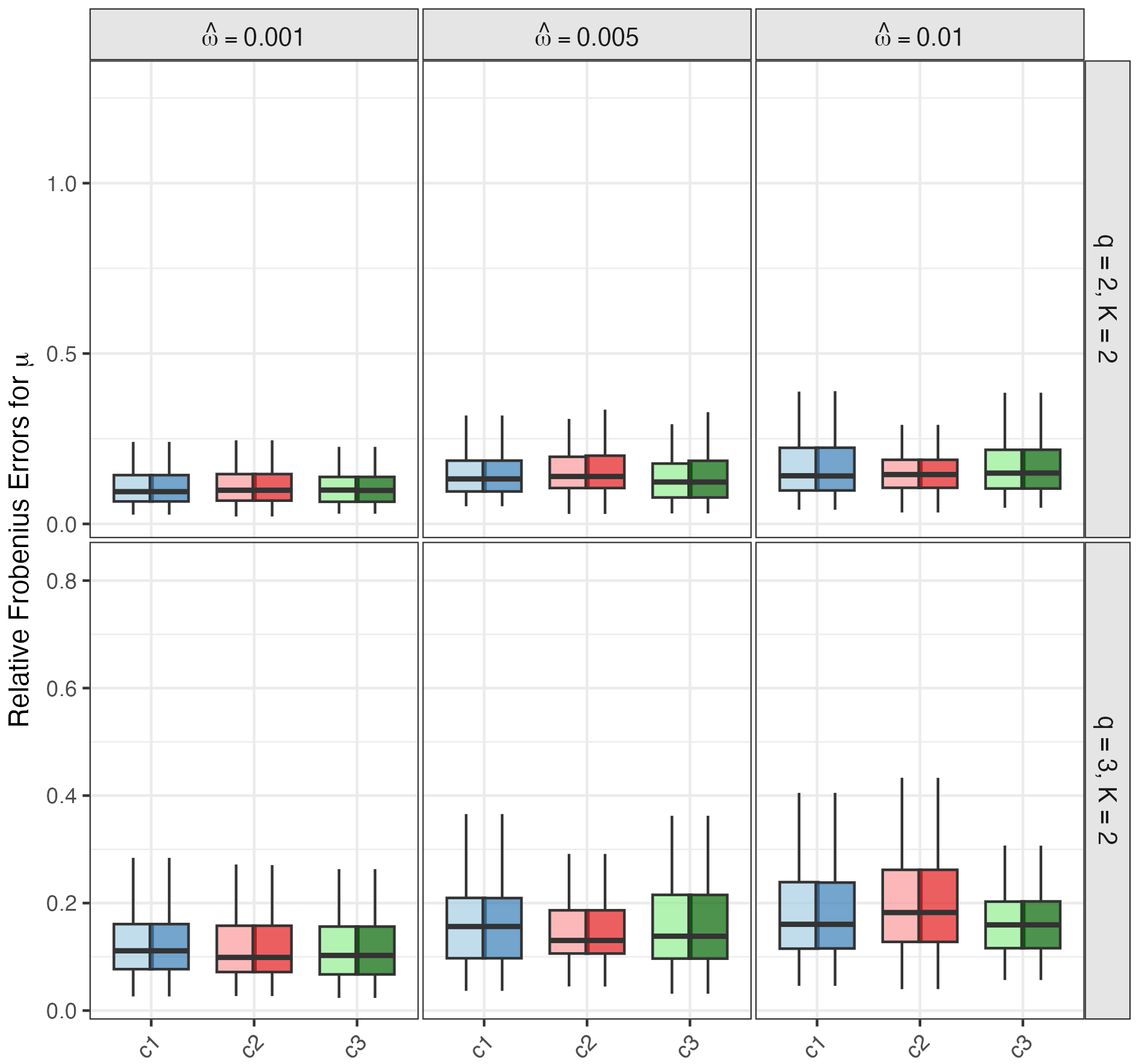}
}
}
\caption{Boxplots of the relative Frobenius errors of estimated parameters $\hat{\Lambda}_k\hat{\Lambda}_k^\top, \hat{\Psi}_k, \hat{\bmu}_k$ for $n=300, p=10$ with varying separation between clusters ($\hat{\omega} = 0.001,0.005,0.01$). Light shades denote results from EMMIX and dark shades denote results from GMMFAD.} 
    \label{fig:ferror-p10}
\end{figure}

\begin{figure}[h!]
    \centering
    \subfloat[$K=2,q = 2,3$]{\includegraphics[width=0.9\linewidth]{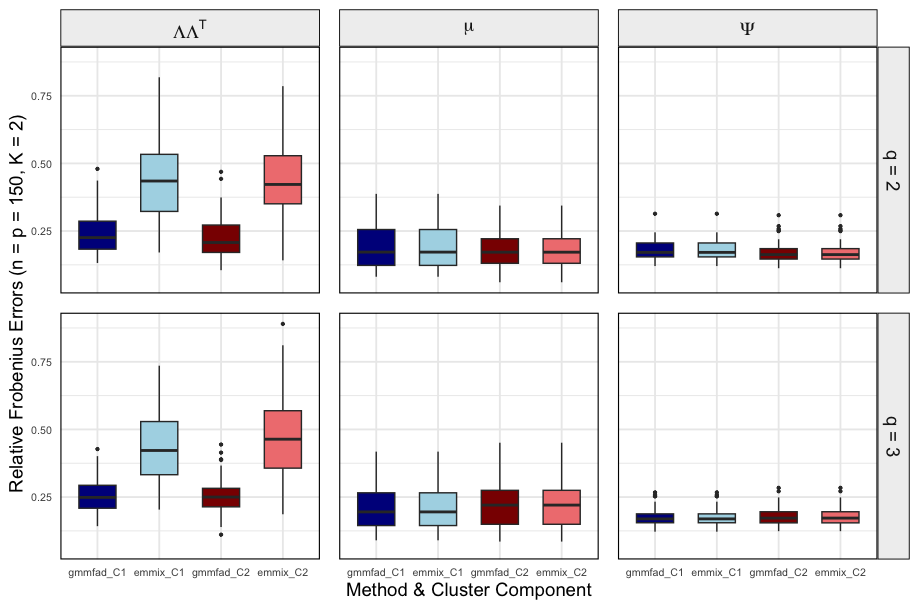}
    }\\
    \subfloat[$K=3,q = 2,3$]{\includegraphics[width=0.9\linewidth]{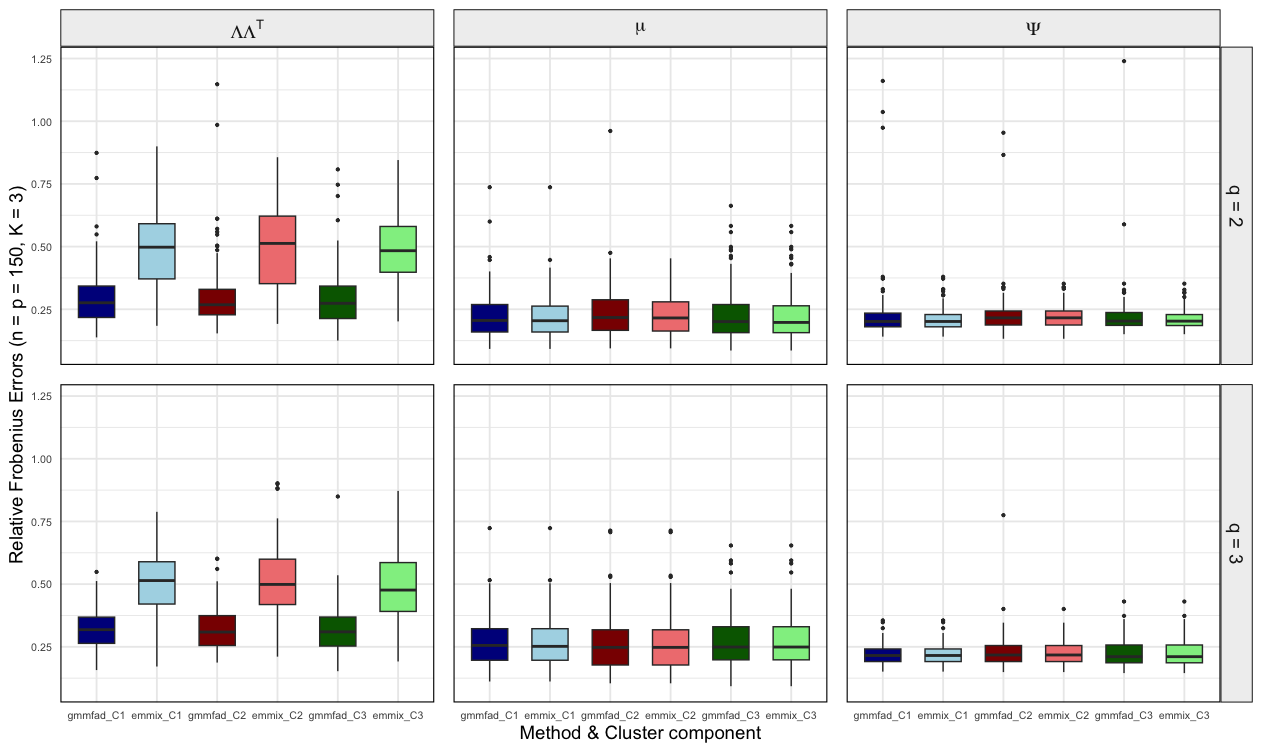}
    }
    \caption{Boxplots of the relative Frobenius errors of parameters $\hat{\Lambda}_k\hat{\Lambda}_k^\top, \hat{\Psi}_k, \hat{\bmu}_k$ for $n=p=150$, with colors indicating different clusters. Light shades denote results from EMMIX and dark shades denote results from GMMFAD. }
    \label{fig:ferror-p150}
\end{figure}
\newpage
\section{Supplementary Figures for Statistical Analysis of Cancer Data}
\label{sec:supp-data}
\begin{figure}[h!]
    \centering
    \includegraphics[width = 0.98\linewidth]{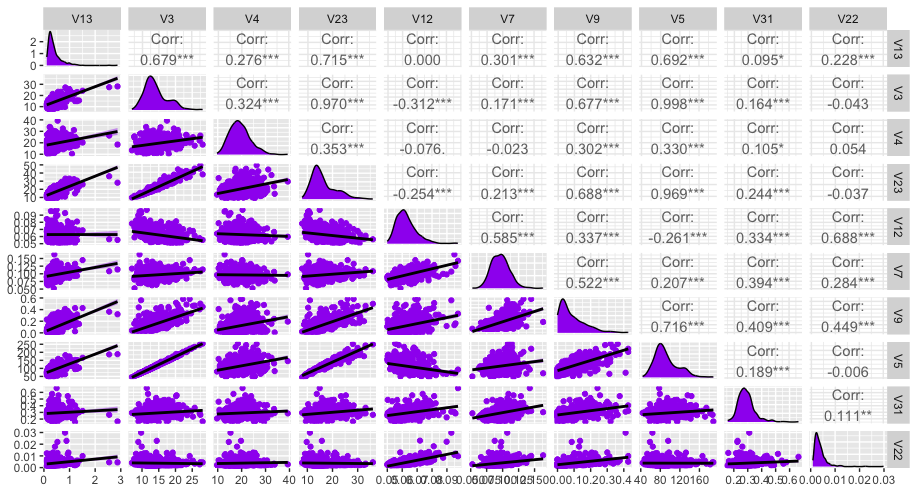}
    \caption{Density and correlation plots of ten randomly selected features of the breast cancer data.}
    \label{fig:wbc_corrplot}
\end{figure}

\begin{figure}[h!]
  \centering
\mbox{
\subfloat[DLBCL]{
\includegraphics[width=0.32\linewidth]{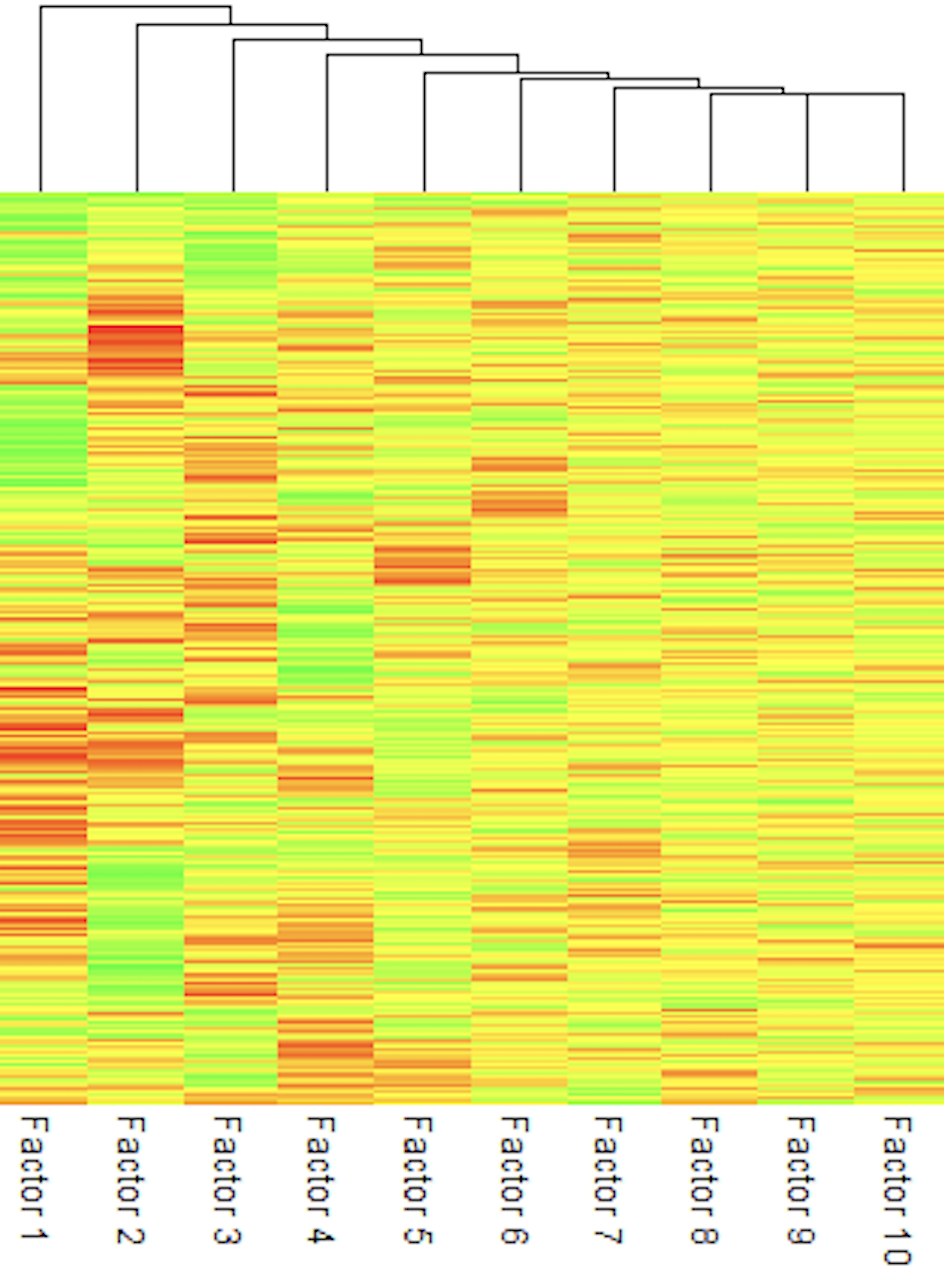}
}
}%
\mbox{
\subfloat[FL]{
\includegraphics[width=0.32\linewidth]{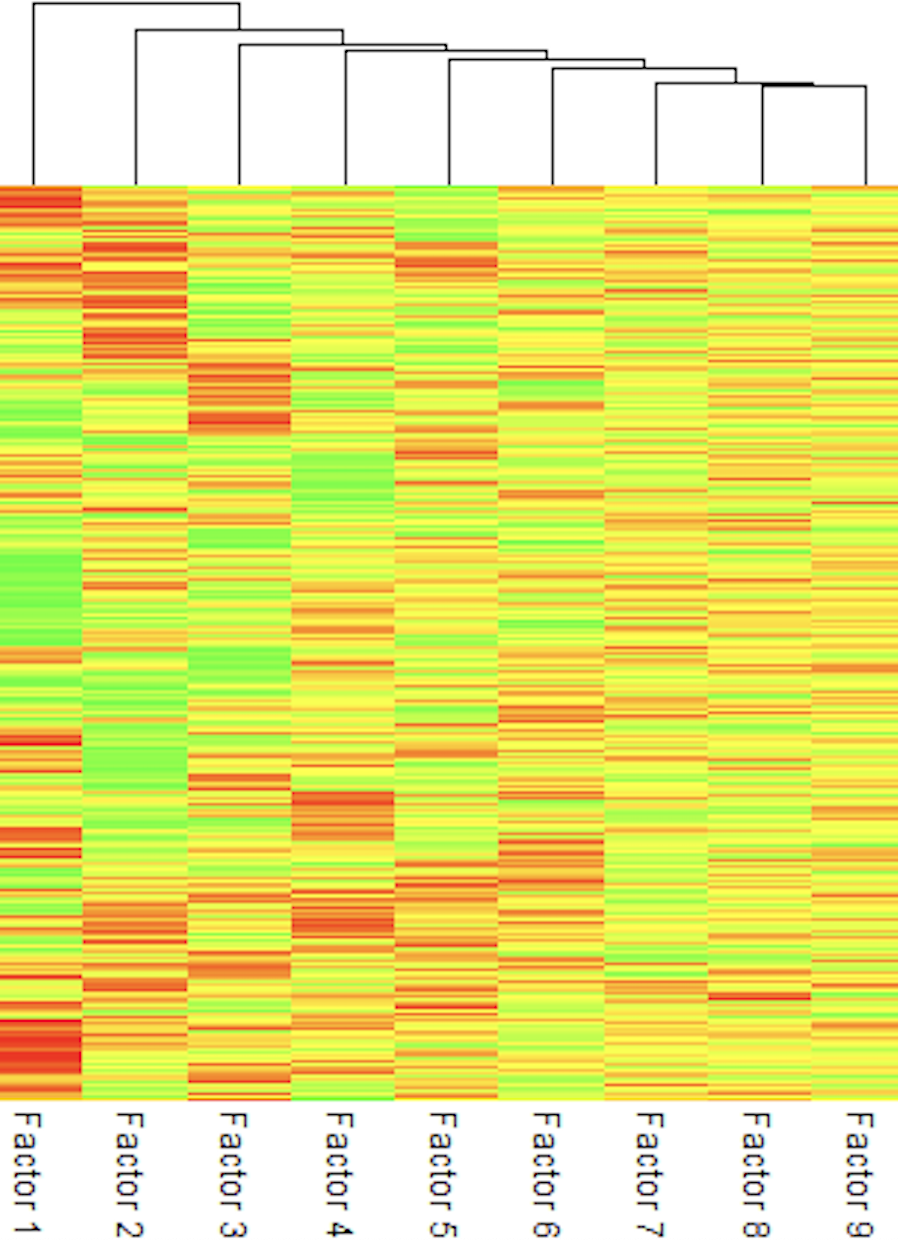}
}
}%
\mbox{
\subfloat[CLL]{
\includegraphics[width=0.32\linewidth]{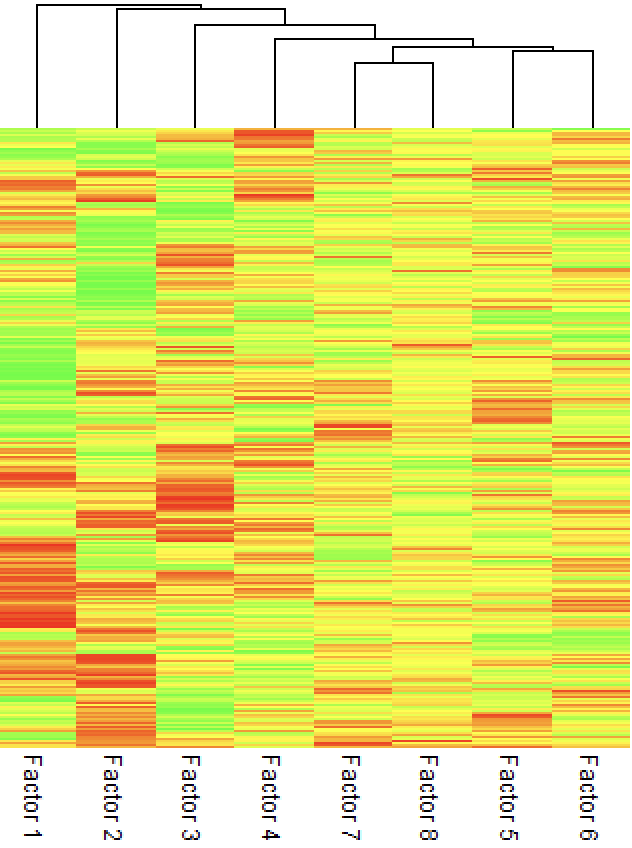}
}
}
\caption{Heatmaps of the estimated factor loadings for the three lymphoma subtypes of (a) DLBCL with ten factors, (b) FL with nine factors, and (c) CLL with eight factors, with colors indicating the loadings values that range from -1 (red) to 1 (green), and dendrogram showing the grouping of factor loadings with similar weights on the $4026$ variables.} 
    \label{fig:loading heatmap_lymphoma}
\end{figure}

\begin{figure}[h!]
  \centering
\mbox{
\subfloat[DLBCL]{
\includegraphics[width=0.675\linewidth]{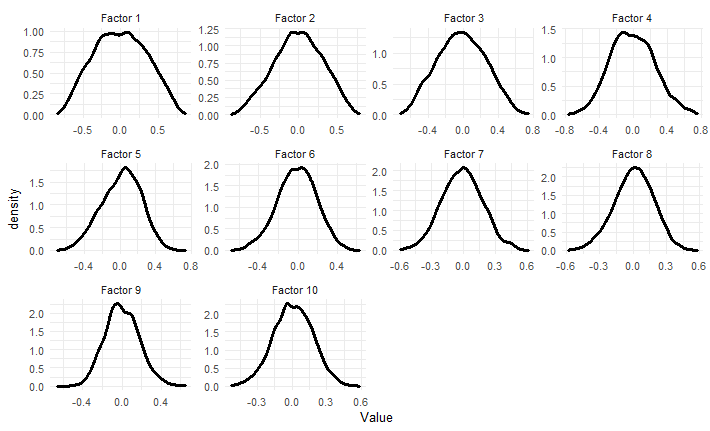}
}
}\\
\vspace{-0.15in}
\mbox{
\subfloat[FL]{
\includegraphics[width=0.675\linewidth]{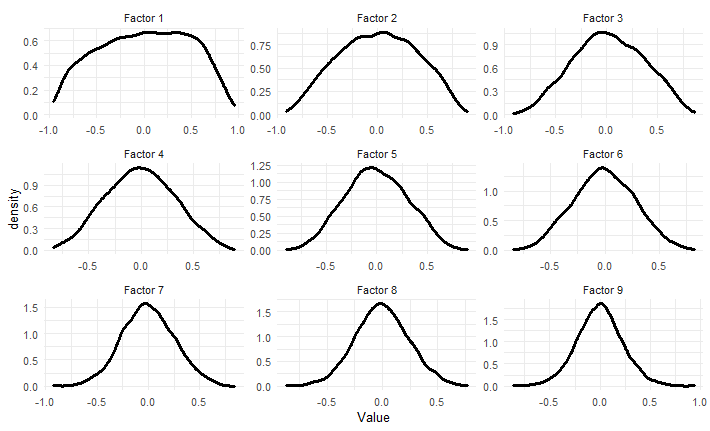}
}
}\\
\vspace{-0.15in}
\mbox{
\subfloat[CLL]{
\includegraphics[width=0.675\linewidth]{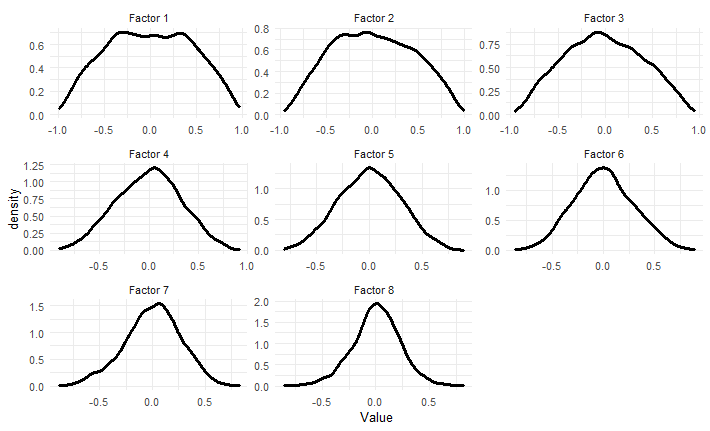}
}
}
\caption{Density curves of the estimated factor loadings for the three lymphoma subtypes of (a) DLBCL with ten factors, (b) FL with nine factors, and (c) CLL with eight factors.}
    \label{fig:loadings distr_curves_lymphoma}
\end{figure}

\end{document}